\documentclass[12pt,lettersize,journal,onecolumn]{IEEEtran}
\usepackage{amsmath,amsfonts,amssymb}
\usepackage{algorithmic}
\usepackage{algorithm}
\usepackage{array}
\usepackage[caption=false,font=normalsize,labelfont=sf,textfont=sf]{subfig}
\usepackage{textcomp}
\usepackage{stfloats}
\usepackage{url}
\usepackage{verbatim}
\usepackage{graphicx}
\usepackage{ntheorem}
\usepackage[algo2e]{algorithm2e} 
\usepackage{amsmath}
\usepackage{cite}
\usepackage{amsfonts}
\usepackage{lipsum}
\usepackage{mathtools}
\usepackage{cuted}
\usepackage{setspace}
\usepackage{multicol}
\usepackage{booktabs, multirow}
\usepackage{siunitx}
\usepackage{tikz}
\usetikzlibrary{shapes, arrows, positioning}
\usepackage{kantlipsum}
\usepackage{footnote}
\usepackage{bm}
\usepackage{amsmath,amsfonts,amssymb}
\usepackage{float}
\usepackage{pifont}     % for \ding{55}
\usepackage{xcolor}     % for colors

\usepackage[sort&compress]{cleveref}

\crefrangeformat{equation}{(\mbox{#3#1#4}-\mbox{#5#2#6})}

\newcommand{\cmark}{\textcolor{green!60!black}{\checkmark}}   % green check
\newcommand{\xmark}{\textcolor{red}{\ding{55}}}               % red cross
\newcommand{\pmark}{\textcolor{orange!80!black}{Partial}}

\newtheorem{pavikl}{\textbf{Lemma}}
\newtheorem{pavikt}{\textbf{Theorem}}

\makeatletter

\newcommand{\Rmnum}[1]{\expandafter\@slowromancap\romannumeral #1@}
\def\BibTeX{{\rm B\kern-.05em{\sc i\kern-.025em b}\kern-.08em
    T\kern-.1667em\lower.7ex\hbox{E}\kern-.125emX}}
\usepackage{balance}
\usepackage{setspace}
\begin{document}
\title{Energy- and Spectral- Efficiency Trade-off in Distributed Massive-MIMO Networks}

\author{\IEEEauthorblockN{Mohd Saif Ali Khan*, Karthik R.M.$^+$, and Samar Agnihotri*}\\
\IEEEauthorblockA{*School of Computing \& EE, Indian Institute of Technology Mandi, HP, India}\\
\IEEEauthorblockA{$^+$Ericsson India Pvt. Ltd., Chennai, TN, India}\\
Email: d21013@students.iitmandi.ac.in, karthik.r.m@ericsson.com, samar@iitmandi.ac.in
%
%\vspace{-0.3in}
}
\maketitle

\begin{abstract}
This paper investigates the energy efficiency (EE) and spectral efficiency (SE) trade-off in uplink distributed massive multiple-input multiple-output (D-mMIMO) systems. Unlike conventional approaches where power consumption focuses primarily on transmit power, we use a comprehensive system-level power consumption framework which incorporates consumption due to fronthaul signaling, distributed processing, and circuit level power, which are, themselves, critically influenced by the dynamic access point (AP) activation (ON/OFF decisions), and AP-user equipment (UE) association strategies. Consequently, we analyze the EE–SE trade-off through the joint optimization of transmit power allocation, AP activation, and AP–UE association. We formulate an optimization problem that maximizes EE while satisfying sum-SE constraints, per-user minimum SE requirements, and fronthaul capacity limits. Our solution uses a fractional programming-based approach to simultaneously determine transmit power levels, dynamic AP-UE associations, and AP activation strategies. Numerical results demonstrate that dynamic AP activation and association substantially impact the EE-SE trade-off, revealing optimal operating points that balance spectral performance with energy consumption. The findings provide practical guidelines for energy-efficient D-mMIMO deployment in next-generation wireless networks, highlighting the importance of adaptive resource allocation in achieving sustainable high-performance communications.
\end{abstract}

\section{Introduction}
\label{intro}
\IEEEPARstart{T}{he} exponential growth in connected devices and the increasing demand for high data rates have placed unprecedented pressure on cellular networks, making both spectral efficiency (SE) and energy efficiency (EE) critical design considerations for 5G and beyond networks. While SE ensures higher throughput and better utilization of spectral resources, maximizing SE mainly often comes at the cost of increased power consumption. This leads to significant challenges in designing communication systems that are both high-performing and energy-efficient. The distributed massive multiple-input multiple-output (D-mMIMO) systems have emerged as a transformative architecture for next-generation networks\cite{ngo2017cell}. By spatially distributing access points (APs) across a service area and jointly serving user equipments (UEs) without traditional cell boundaries, D-mMIMO systems improve macro-diversity, mitigate cell-edge issues, and enable uniform service quality \cite{ngo2017cell, khan2024distributed}. Within this architecture, three system-level factor, power allocation, AP-UE association and AP switching (ON/OFF strategies), play a major role in determining both EE and SE \cite{ngo2017cell,buzzi2017cell,chen2022sparse,ngo2017total}. 

The AP-UE association is particularly crucial as it directly governs both the achievable SE through interference management and cooperative gain, as well as the energy consumption via fronthaul signaling and distributed processing overhead. Similarly, the AP activation serves as a fundamental lever in the EE-SE trade-off, where the number of simultaneous active APs impacts the total power consumption (circuit, fronthaul, and processing costs) and the maximum attainable system throughput. Unlike centralized MIMO systems where transmit power and base station operating costs dominate total energy consumption, D-mMIMO introduces significant additional energy costs due to distributed circuit operations, fronthaul communication, and cooperative signal processing \cite{chen2022sparse,ngo2017total,chen2023energy}. Consequently, the number of active APs and their user associations significantly impact system-wide energy consumption, with these infrastructure costs often dominating the total transmission energy costs in the total energy consumption. From a deployment perspective, understanding the EE-SE trade-off is crucial for network planning and operational decisions. Network operators face fundamental questions: How many APs should be activated to meet specific throughput requirements? What association patterns maximize energy savings without compromising service quality? How should resources be adapted under varying traffic loads? Existing approaches that optimize EE or SE in isolation provide limited solution guidance for such practical trade-offs. 

\subsection{Related Works}
\subsubsection*{Centralized MIMO and Early Distributed Approaches}
The EE-SE trade-off has been extensively studied for centralized MIMO systems \cite{ngo2013energy,huang2018spectral,yang2021millimeter}, where optimization primarily focuses on transmit power consumption. However, these approaches are inadequate for distributed architectures as they neglect the significant energy costs of distributed processing and fronthaul links. Early distributed antenna systems \cite{he2012energy,onireti2013energy,jiang2013relation} acknowledged backhaul power but treated distributed antenna units' power as constants, optimizing only transmit power. Similarly, \cite{yang2018energy} considered EE in D-mMIMO but used transmit-power-only models, optimized via signal-to-interference-noise (SINR) without explicit EE-SE trade-off consideration.

\subsubsection*{EE-SE trade-off with Realistic Power Models but Static Configurations}
Recent studies have incorporated more realistic power consumption models. Works like \cite{chen2022sparse,chen2023energy,huang2024performance} include comprehensive power models encompassing fronthaul and processing costs while analysing EE-SE trade-off. However, these approaches typically employ fixed AP-UE associations and assume that all APs remain active, failing to utilized these critical degrees of freedom.  In \cite{bashar2019energy,alonzo2019energy}, EE maximization considers only transmission power without dynamic AP-UE association and AP activation. 

\subsubsection*{Partial Optimization Approaches}
Several works have addressed aspects of dynamic resource allocation but fall short of comprehensive EE-SE trade-off analysis. The authors in \cite{ngo2017total} consider AP-UE association and power allocation for downlink EE maximization but without analyzing the fundamental EE-SE trade-off. The authors in \cite{vu2020joint} present joint power control and active AP selection for downlink but assume that all active APs serve all users, negating dynamic association benefits. A recent work \cite{hong2025energy} considers EE related to AP switching but in isolation from power control and AP-UE association variables. Even a machine learning approach in \cite{ooi2024joint} proposes dynamic AP sleep modes, but assumes fixed serving patterns and minimize total energy without explicit EE-SE trade-off optimization.

\subsubsection*{SE-Focused Optimization with Neglected EE Implications}
Many studies pursue SE maximization or power minimization while overlooking energy efficiency implications. Works like \cite{ngo2018performance,guenach2020joint,bjornson2020scalable} optimize uplink SE through AP selection and power allocation without EE considerations. The authors in \cite{mai2022energy} address EE maximization but focus only on transmit power optimization. Recent studies \cite{hao2024joint,khan2024joint} consider joint AP-UE association and power allocation for SE maximization but overlook EE. 

\subsubsection*{Comprehensive EE-SE Trade-off but Impractical Approaches}
The work in \cite{demir2024cell} represents the most comprehensive downlink approach, formulating optimization problems for power allocation, AP-UE association, and AP activation. However, critical limitations remain: the trade-off problem neglects user fairness, and the use of local partial MMSE (LP-MMSE) precoding requires Monte-Carlo expectation calculations, making the optimization dependent on small-scale fading. This necessitates AP switching decisions every few milliseconds, depending upon the length of coherence block, making it highly impractical for real-world deployment. Similarly in  \cite{nguyen2020spectral}, the authors, for uplink, assume that all UEs are served by all active APs, limiting optimization to transmit power and AP activation without the  AP-UE association. They have considered a perfect CSI case during trade-off optimization, thus making the optimization dependent on small-scale fading, and impractical to implement.

\begin{table*}[htbp]
\centering
\caption{Comparison of Related Works on D-mMIMO Resource Optimization}
\label{tab:literature_comparison}
\renewcommand{\arraystretch}{0.8} 
\resizebox{0.95\textwidth}{!}{%
\begin{tabular}{|l|c|c|c|c|c|c|c|c|}
\hline
\shortstack[c]{\rule{0pt}{3ex}\textbf{Work}} & 
\shortstack[c]{\rule{0pt}{3ex}\textbf{Power} \\ \textbf{Control}} & 
\shortstack[c]{\rule{0pt}{3ex}\textbf{AP-UE} \\ \textbf{Association}}& 
\shortstack[c]{\rule{0pt}{3ex}\textbf{AP} \\ \textbf{Switching}} & 
\shortstack[c]{\rule{0pt}{3ex}\textbf{Capacity} \\ \textbf{specific} \\ \textbf{Trade-off}} & 
\shortstack[c]{\rule{0pt}{3ex}\textbf{Uplink}} & 
\shortstack[c]{\rule{0pt}{3ex}\textbf{Realistic}\\ \textbf{Power}} & 
\shortstack[c]{\rule{0pt}{3ex}\textbf{User} \\ \textbf{Fairness}} & 
\shortstack[c]{\rule{0pt}{3ex}\textbf{LSFCs} \\ \textbf{Timescale}} \\
\hline
Ngo et al. \cite{ngo2017total}        & \cmark & \cmark & \xmark & \xmark & \xmark & \cmark & \cmark & \cmark \\
\hline
Chen et al. \cite{chen2022sparse,chen2023energy} & \xmark & \xmark & \xmark & \xmark & \cmark & \cmark & \xmark & \cmark \\
\hline
Yang et al. \cite{yang2018energy}     & \cmark & \xmark & \xmark & \xmark & \xmark & \xmark & \cmark & \cmark \\
\hline
Bashar et al. \cite{bashar2019energy} & \cmark & \xmark & \xmark & \xmark & \cmark & \xmark & \cmark & \cmark \\
\hline
Alonzo et al. \cite{alonzo2019energy} & \cmark & \xmark & \xmark & \xmark & \cmark & \xmark & \xmark & \xmark \\
\hline
Vu et al. \cite{vu2020joint}          & \cmark & \xmark & \cmark & \xmark & \xmark & \cmark & \cmark & \cmark \\
\hline
Hong et al. \cite{hong2025energy}     & \xmark & \xmark & \cmark & \xmark & \xmark & \cmark & \xmark & \cmark\\
\hline
Mai et al. \cite{mai2022energy}       & \cmark & \xmark & \xmark & \xmark & \xmark & \cmark & \cmark & \cmark \\
\hline
Ooi et al. \cite{ooi2024joint}        & \cmark & \xmark & \cmark & \xmark & \cmark & \xmark & \cmark & \cmark \\
\hline
Demir et al. \cite{demir2024cell}     & \cmark & \cmark & \cmark & \pmark & \xmark & \cmark & \xmark & \xmark \\
\hline
Nguyen et al. \cite{nguyen2020spectral} & \cmark & \xmark & \cmark & \pmark & \cmark & \cmark & \cmark & \xmark \\
\hline
\textbf{Our Work}                     & \cmark & \cmark & \cmark & \cmark & \cmark & \cmark & \cmark & \cmark \\
\hline
\end{tabular}%
}
\end{table*}

\subsubsection*{Critical Research Gaps}
The literature survey reveals (Table \ref{tab:literature_comparison}) limitations across four dimensions:
\begin{itemize}
\item \textbf{Incomplete Power Modeling}: Most works optimize transmit power while neglecting the substantial impact of the AP activation and the AP-UE association on fronthaul and processing power.
\item \textbf{Downlink Bias}: Most comprehensive joint optimization methods target downlink scenarios where transmit power and the AP–UE association are inherently coupled which means an AP allocates non-zero power only to the UEs it serves, and zero otherwise. However, uplink optimization presents distinct challenges as transmit power is broadcast irrespective of AP-UE association, requiring fundamentally different approaches.
\item \textbf{Methodological Limitation in EE-SE Problem Formulation:} Existing works predominantly use weighted-sum approaches between EE and SE or bi-objective optimization of EE and SE with abstract trade-off factors that lack clear operational interpretation. This contrasts with practical network operation where operators need to maximize EE while meeting specific throughput targets, not balance abstract weights corresponding to bi-objective optimization of EE and SE.
\item \textbf{Practical Implementation Barriers}: Even theoretically comprehensive approaches often rely on impractical assumptions, such as frequent re-optimization every coherence block or neglect of essential constraints like minimum per user SE.
\end{itemize}
Unlike prior works that formulate EE–SE trade-offs using abstract weighted objectives or rely on instantaneous CSI-dependent optimization, this work maximizes EE under explicit throughput constraints using only large-scale fading statistics, enabling optimization on network-control timescales rather than coherence-block timescales. This distinction is particularly important in uplink D-mMIMO systems, where transmit power is decoupled from AP–UE association, receiver-side combining replaces precoding, and decoding-related infrastructure power depends on long-term association and activation decisions. These characteristics fundamentally alter the EE–SE trade-off structure compared to downlink formulations and motivate a joint uplink design that accounts for power control, AP–UE association, and AP activation under realistic power consumption models.

\subsection{Contributions}
Motivated by the limitations of existing EE–SE optimization approaches in uplink D-mMIMO systems, this work develops a practical framework for energy-efficient uplink operation under explicit throughput and fairness constraints. The main contributions of this paper, with emphasis on practical uplink D-mMIMO deployment, are summarized as follows:

\begin{itemize}
\item \textbf{Capacity-Constrained EE-SE Framework for Uplink D-mMIMO:}
We formulate an uplink energy-efficiency maximization framework under explicit sum SE and per-user QoS constraints, using only large-scale fading statistics. Unlike weighted multi-objective EE–SE formulations or CSI-dependent optimization approaches, the proposed formulation aligns with practical network operation by enabling optimization on network-control timescales rather than coherence-block timescales, while accurately capturing the impact of AP activation and AP–UE association on infrastructure power consumption.

\item \textbf{Joint Optimization of Power Allocation, AP-UE Association, and AP Switching:}
We formulate and solve a joint optimization problem by using the fractional programming with nested and simple quadratic transform that simultaneously determines transmit power levels, dynamic AP-UE associations, and the AP activation strategies to maximize overall energy efficiency while satisfying sum-SE constraints. This joint optimization enables adaptive network reconfiguration based on throughput requirements or energy priorities, achieving significant performance gains over isolated optimization approaches. Our optimization framework also incorporates critical practical considerations including per-user minimum SE requirements to ensure service quality, and fronthaul load control through a constraint on the maximum number of UEs served per AP. 

\item \textbf{Deployment-Oriented Operational Insights:}
This work provides actionable guidance for network planning and operation by revealing how optimal AP density and association patterns depend on user load and SE targets. We demonstrate that strategic AP  activating/deactivating based on EE-SE trade-off analysis, enables substantial energy savings without compromising throughput requirements, offering operators concrete strategies for sustainable network deployment.

\item \textbf{Numerical Simulations:} Extensive numerical simulations validate the framework under diverse scenarios, demonstrating robust performance, rapid convergence, and significant improvements over baseline approaches across various network configurations.
\end{itemize}
To the best of knowledge, this comprehensive contribution advances the state-of-the-art in D-mMIMO resource allocation by providing both theoretical insights and practical deployment guidelines for energy-efficient next-generation wireless networks.

\textit{Organization:}
The rest of the paper is structured as follows. Section \ref{system_model} presents the system model. Section \ref{problem_formulation} describes the proposed optimization problem. Section \ref{solution_analysis} details the proposed solution methodology. Section \ref{simulations} presents numerical results that analyze the EE-SE trade-off under various scenarios. Section \ref{conclusion} concludes the paper and discusses future directions.

\textit{Notation:}
Scalars are denoted by italic letters (e.g., $x$), vectors by bold lowercase letters (e.g., $\mathbf{x}$), and matrices by bold uppercase letters (e.g., $\mathbf{X}$). The transpose of a matrix or vector is denoted by $(\cdot)^T$, and the Hermitian operation is denoted by $(\cdot)^H$. The complex conjugate of a scalar is denoted by $(\cdot)^*$. The notation $\mathbb{R}$ and $\mathbb{C}$ represent the sets of real and complex numbers, respectively.  The expectation operator is denoted by $\mathbb{E}[\cdot]$. The identity matrix of size $N$ is denoted by $\mathbf{I}_N$. A complex Gaussian random vector $\mathbf{x}$ with mean $\boldsymbol{\mu}$ and covariance $\mathbf{K}$ is denoted as $\mathbf{x} \sim \mathcal{CN}(\boldsymbol{\mu}, \mathbf{K})$. A diagonal matrix of $\mathbf{x}$ is represented as $diag(\mathbf{x})$. The big-O notation is represented by $\mathcal{O}$.

\section{System Model}
\label{system_model}
\begin{figure}[!t] 
\centering 
\includegraphics[width=0.40\textwidth]{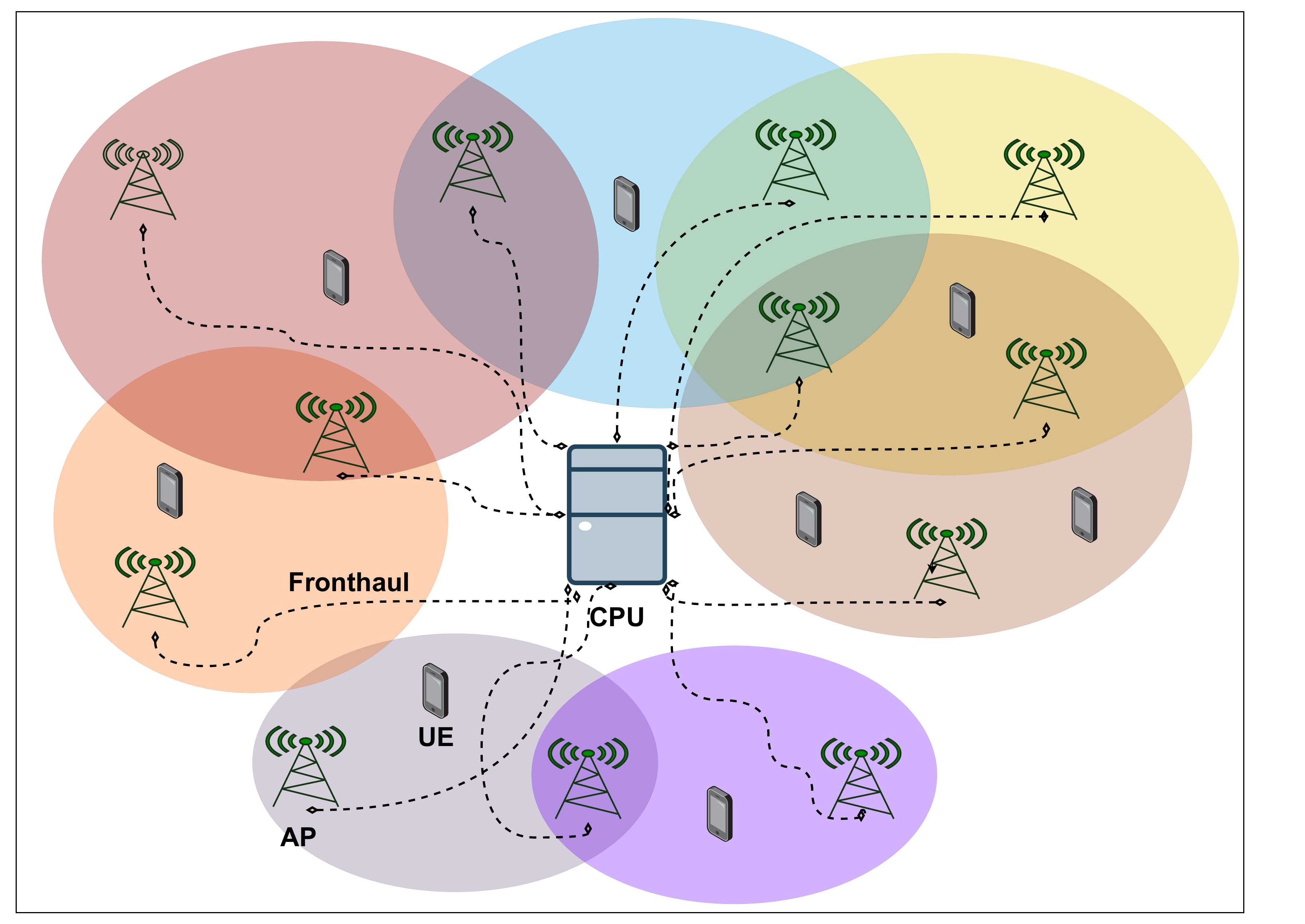} 
\caption{Illustration of a user-centric distributed massive MIMO system. Each oval indicates a UE and the APs serving it.} 
\label{fig_4} 
\end{figure}

We consider the uplink of a D-mMIMO system comprising $T$ single-antenna UEs and $M$ APs, each equipped with $A$ antennas, uniformly deployed over a defined coverage area. Each AP is connected to a centralized processing unit (CPU) via reliable fronthaul link, enabling coordinated processing, data exchange, and user scheduling across the network. The system operates under a user-centric transmission paradigm, where each UE is served by a subset of geographically proximate APs, as illustrated in Fig.~\ref{fig_4}. This approach enhances scalability and reduces fronthaul signaling load compared to fully connected cell-free architectures.
The system employs a block fading model, where the channel remains constant over a coherence interval of $L_c$ symbols and changes independently across blocks. Each coherence block dedicates $L_p$ symbols for uplink pilot transmission. All APs operate over the same time-frequency resources and are capable of simultaneously serving multiple users. Let $\mathbf{h}_{mt} {\in} \mathbb{C}^{A \times 1}$ denote the small-scale fading vector between the $m$-th AP and the $t$-th UE. We assume independent Rayleigh fading, such that $\mathbf{h}_{mt} {\sim} \mathcal{CN}(0, \mathbf{I}_A)$, where all components are i.i.d. complex Gaussian with zero mean and unit variance. The large-scale fading coefficient (LSFC) $\beta_{mt}$ accounts for path loss and shadowing. Thus, the overall channel vector is $\textbf{g}_{mt} {=} \beta_{mt}^{1/2} \textbf{h}_{mt}$.

\subsection{Uplink Pilot Training}
During training, each UE \( t {\in} \{1, {\dots}, T\} \) transmits a pilot sequence \( \sqrt{L_p}\boldsymbol{\psi}_{i_t} {\in} \mathbb{C}^{L_p \times 1} \), satisfying \( \|\boldsymbol{\psi}_{i_t}\|^2 {=} 1 \). The received pilot signal at AP \( m \) is expressed as:
\begin{align*}
\textbf{y}^{\mathrm{pilot}}_m = \sum\nolimits_{t=1}^T \sqrt{L_p p_p} \, \mathbf{g}_{mt} \boldsymbol{\psi}_{i_t}^\mathsf{T} + \mathbf{N}_m,
\end{align*}
where \( \mathbf{N}_m {\in} \mathbb{C}^{A \times L_p} \) denotes the additive white Gaussian noise matrix with i.i.d. \( \mathcal{CN}(0,\sigma^2) \) entries, and \( p_p \) is the uplink maximum pilot power. The MMSE estimate of channel \( \mathbf{g}_{mt} \),   at AP \( m \) is given by~\cite{ngo2017cell}:
\begin{align*}
\hat{\textbf{g}}_{mt} = \sqrt{p_pL_p}\beta_{mt}\theta^{-1}_{mi_t}\textbf{y}^{pilot}_{m}\boldsymbol{\psi}_{i_t},
\end{align*}
where $ \theta_{mi_t} {=}\big(\sum_{k=1}^{T}p_pL_p\beta_{mk}\left|\boldsymbol{\psi}^{H}_{i_t}\boldsymbol{\psi}_{i_k} \right|^2 {+} \sigma^2\big)$.
The estimate $\hat{\mathbf{g}}_{mt}$ and estimated error $\tilde{\mathbf{g}}_{mt}$ are independent Gaussian with distributions $\hat{\mathbf{g}}_{mt} {\sim} \mathcal{CN}(0, \gamma_{mt}\textbf{I}_A)$ and $\tilde{\mathbf{g}}_{mt} {\sim} \mathcal{CN}(0, (\beta_{mt}{-}\gamma_{mt})\textbf{I}_A)$, where
\begin{align*}
\gamma_{mt} = p_pL_p\beta_{mt}^2\theta^{-1}_{mi_t}.
\end{align*}

\subsection{Uplink Data Transmission}
During the data transmission phase, the received uplink signal at AP \( m \) is:
\begin{align}
\label{etr_1}
\mathbf{y}^{\mathrm{ul}}_m =  \sum\nolimits_{t=1}^T \mathbf{g}_{mt} \sqrt{p_u\eta^u_t} x_t + \mathbf{n}_m,
\end{align}
where \( x_t \) is the data symbol transmitted by  UE \( t \), satisfying \( \mathbb{E}\{|x_t|^2\} {=} 1 \); \( \eta^u_t {\in} [0,1] \) is the uplink power control coefficient for UE \( t \); and \( p_u \) denotes the uplink maximum power. The noise vector \( \mathbf{n}_m {\in} \mathbb{C}^{A \times 1} \) contains i.i.d. \( \mathcal{CN}(0,\sigma^2) \) entries. Each AP applies a local combining vector \( \mathbf{v}_{mt} {\in} \mathbb{C}^{A \times 1} \) to detect the signal of UE \( t \). The partial detection \( \hat{x}_{mt} {=} \mathbf{v}_{mt}^\mathsf{H} \mathbf{y}^{\mathrm{ul}}_m \) is forwarded to the CPU. The CPU performs Large-Scale Fading Decoding (LSFD) across the APs~\cite{chen2022sparse,zhang2021local}:
\begin{align}
\label{etr_2}
\hat{x}_t = \sum\nolimits_{m=1}^{M} d_{mt}a_{mt} \hat{y}_{mt} = \sum\nolimits_{m=1}^{M} d_{mt} a_{mt} \mathbf{v}_{mt}^\mathsf{H} \mathbf{y}^{\mathrm{ul}}_m,
\end{align}
where \( d_{mt} {\in} \{0,1\} \) is the AP-UE association indicator: \( d_{mt} {=} 1 \) if AP \( m \) serves UE \( t \); otherwise \( d_{mt} {=} 0 \). Also, $a_{mt}$ is the LSFD coefficient for UE $t$ with respect to the $m$. Substituting \( \mathbf{y}^{\mathrm{ul}}_m \) in \eqref{etr_2}, the overall estimated signal becomes~\cite{zhang2021local}:
\begin{align*}
%\label{etr_3}
\begin{split}
\hat{x}_{t}   & = \!\! \sum_{m=1}^{M}\!\!\!\sqrt{p_u\eta^u_t}d_{mt}a_{mt}\mathbf{v}_{m{t}}^{H}\hat{\textbf{g}}_{mt}x_t  + \!\! \sum_{m=1}^{M}\!\!\!\sqrt{p_u\eta^u_t}d_{mt}a_{mt}\mathbf{v}_{m{t}}^{H}\tilde{\textbf{g}}_{mt}x_t + \! \sum_{\substack{k=1 \\ k \neq t}}^T  \sum_{m=1}^{M}\!\!\!\sqrt{p_u\eta^u_k}d_{mt}a_{mt}\mathbf{v}_{m{t}}^{H}\textbf{g}_{mk}x_k  \\ & + \!\! \sum_{m=1}^{M}\!\!d_{mt}a_{mt}\mathbf{v}_{m{t}}^{H}\textbf{n}_{m}.
\end{split}
\end{align*}
Here, the first term is the desired signal for UE $t$, the second term is interference term due to imperfect channel estimation, the third term is interference term due to all other UEs and the last term is the noise. 

We have considered Partial Full-Pilot Zero-Forcing (PFZF) combining~\cite{zhang2021local}, where each AP suppresses the interference caused by UEs with strong channels for other UEs who also have strong channels, using local zero-forcing vectors, provided the AP has sufficient spatial degrees of freedom. Specifically, each AP \( m \) classifies the users it serves into two groups: the strong users, denoted by \( \mathcal{S}_m \), whose interference is actively suppressed, and the weak users, denoted by \( \mathcal{W}_m \), whose interference is only partially mitigated due to limited spatial resources. In addition, we also define a binary variable $\delta_{mt}$ such that $\delta_{mt} {=}1$, when $t {\in} \mathcal{S}_m$ and $\delta_{mt} {=} 0$, otherwise. Let $L_{\mathcal{S}_m}$ denote the number of pilots used by strong UEs at AP $m$. The partitioning of UEs into strong or weak are performed in the same way as in \cite{zhang2021local}, while maintaining $A {-} L_{\mathcal{S}_m}{>}0$. Although further performance gains can be achieved using generalized PFZF (G-PFZF)~\cite{khan2025adaptive}, this work focuses on the standard PFZF scheme.
The uplink SE of UE \( t \) is lower bounded as~\cite{zhang2021local,khan2025comments}:
\begin{align}
\label{eq:SE_1}
\mathrm{SE}^{u}_{t} = w \log_2\left(1 + {\Gamma_t} \right),
\end{align}
where \( w = \frac{\big(1 - \frac{L_p}{L_c} \big)}{2} \) is the pre-log factor for uplink transmission and $\Gamma_t$ is the SINR for UE $t$ is given by
\begin{align}
\label{eq:SINR_1}
\Gamma_t =  \frac{\mathrm{DS}_t}{\mathrm{PC}_t + \mathrm{BU}_t + \mathrm{NI}_t + \mathrm{N}_t}.
\end{align}
The term \( \mathrm{DS}_t \) in \eqref{eq:SINR_1} represents the desired signal component, given by~\cite{zhang2021local,khan2025comments}
\begin{align}
\label{eq:SE_2}
\mathrm{DS}_t = {p_u \eta^u_t} \Big| \sum\nolimits_{m =1}^M d_{mt}a^*_{mt}\sqrt{(A-L_{\mathcal{S}_m}) \gamma_{mt}} \Big|^2.
\end{align}
 The pilot contamination term \( \mathrm{PC}_t \) accounts for coherent interference from users sharing the same pilot and is expressed as~\cite{zhang2021local,khan2025comments}
\begin{align}
\label{eq:SE_3}
\mathrm{PC}_t = \!\!\!\sum_{ k\in \mathcal{P}_t/\{t\}}\!\!\!\! \eta^u_{k}p_u\Big| \sum\nolimits_{m =1}^M d_{mt}a^*_{mt}\sqrt{(A-L_{\mathcal{S}_m}) \gamma_{mk}} \Big|^2.
\end{align}
Here, \( \mathcal{P}_t \) denotes the set of UEs sharing the same pilot sequence as UE \( t \). The term \( \mathrm{BU}_t \) captures the beamforming uncertainty due to the mismatch between the actual channel and its estimate as~\cite{zhang2021local,khan2025comments}
\begin{align}
\label{eq:SE_4}
\mathrm{BU}_t = p_u \eta^u_t 
\sum\nolimits_{m =1}^M d_{mt}|a^*_{mt}|^2 (\beta_{mt} - \delta_{mt}\gamma_{mt}),
\end{align}
 The non-coherent interference from other interfering users is represented by \( \mathrm{NI}_t \), given by~\cite{zhang2021local,khan2025comments}
\begin{align}
\label{eq:SE_5}
\mathrm{NI}_t =\!\! \sum_{k \neq t} p_u \eta^u_{k} 
\sum\nolimits_{m =1}^M d_{mt}|a^*_{mt}|^2 (\beta_{mk} - \delta_{mt}\delta_{mk}\gamma_{mk}),
\end{align}
Finally, \( \mathrm{N}_t \), accounts for the noise, is given as~\cite{zhang2021local,khan2025comments}
\begin{align}
\label{eq:SE_6}
\mathrm{N}_t = \sigma^2\sum\nolimits_{m =1}^M d_{mt} |a^*_{mt}|^2.
\end{align}
The LSFD vector $\textbf{a}_t = \{a_{mt}\}_{m {\in} \mathcal{M}}$ for UE $t$ is given by~\cite{zhang2021local,khan2025comments}
\begin{align}
\label{eq:LSFD}
\textbf{a}_t = \mathbf{C}^{-1}_t\textbf{b}_{tt},\ \ \text{where},
\end{align}
\begin{align}
\label{eq:LSFD_1}
\mathbf{C}_{t} &=\!\!\!\! \sum\nolimits_{k \in \mathcal{P}_t/\{t\}}\! p_u \eta^u_{k}\mathbf{b}_{kt}\mathbf{b}^H_{kt} + diag({W}_{1t},{W}_{2t},..,{W}_{Mt}),
\end{align}
\begin{align}
\label{eq:LSFD_2}
{W}_{mt} =&  \sum\nolimits_{k=1}^{T} p_{u}\eta^u_k \left( \beta_{mk} - \delta_{mt} \delta_{mk} \gamma_{mk} \right) + \sigma^2
\end{align}
\begin{align}
\label{eq:LSFD_3}
[\mathbf{b}_{kt}]_m =  \sqrt{(A - \delta_{mt}L_{\mathcal{S}_m})\, \gamma_{mk}},
\end{align} 

The total uplink energy efficiency of the network (in bits per joule) is defined as:
\begin{align}
\label{eq:EE}
\text{EE} = \frac{wB\sum_{t=1}^{T} \log_2\left(1 + \Gamma_t \right)}{P_{\text{T}}},
\end{align}
where \( B \) is the total bandwidth and, following \cite{chen2022sparse}, the total power is given by
\begin{align*}
P_{\text{T}} &= P^{\text{fix}}_{\text{T}} + P^c_{\text{T}}  + wP^{\text{deco}}_{cpu}B\sum_{t=1}^T\log_2\left(1 + \Gamma_t \right), \\
P^c_{\text{T}} &= \sum_{t=1}^{T} \Bigg( \frac{\eta^u_t p_u}{\zeta} + \sum_{m=1}^{M} d_{mt} P^{\text{lsfd}}_{cpu}  +   A d_{mt} P^{\text{proc}} + d_{mt} P^{\text{sig}} \Bigg), \\
P^{\text{fix}}_{\text{T}} &= T P^{\text{c}}_{ue} + \sum_{m=1}^{M} s_m\Big(A P^{\text{c}}_{ap} +  P^{\text{fix}}_{fh}\Big) + P^{\text{fix}}_{cpu}.
\end{align*}
Here, \( P^{\text{c}}_{ue} \), \( P^{\text{c}}_{ap} \), \( P^{\text{proc}} \), \( P^{\text{fix}}_{fh} \), \( P^{\text{sig}} \), \( P^{\text{fix}}_{cpu} \), \( P^{\text{lsfd}}_{cpu} \), and \( P^{\text{deco}}_{cpu} \) denote the circuit power of UEs and APs, signal processing power, fronthaul fixed power, fronthaul processing power, fixed CPU power and CPU processing powers for LSFD and decoding, respectively. The parameter \( \zeta \in (0,1] \) denotes the power amplifier efficiency. Also, $s_m$ is the binary variable related to the AP activation, such that \( s_{m} {=} 1 \) if AP \( m \) is active; otherwise \( s_{m} {=} 0 \). Crucially, the binary variables $s_m$  and $d_{mt}$  directly control substantial energy costs: when $s_m = 0$, the AP's entire circuit power, fronthaul connectivity, and processing overhead are eliminated, while each active association $d_{mt} = 1$ adds processing and signaling energy. In dense deployments, these infrastructure components often dominate total energy consumption, making AP activation and association decisions more impactful than transmit power alone.

\section{Problem Formulation}  
\label{problem_formulation}

In this section, we formulate a comprehensive optimization problem that maximizes the energy efficiency by jointly optimizing three critical decision variables: power allocation coefficients $\bm{\eta^{u}}$, AP-UE association matrix $\textbf{D}$, and AP activation vector $\textbf{s}$. The optimization is subject to meeting both system-level sum spectral efficiency requirements and per-user minimum SE constraints, while also incorporating practical fronthaul limitations. The problem is expressed as:
\begin{subequations}
\label{eq_4_m}
\begin{align} 
\label{eq_4}
\underset{\bm{\eta^{u}},\textbf{D},\textbf{s}}{\max~} &  \frac{wB\sum_{t=1}^{T}\log_2(1+\Gamma_t)}{P^{\text{fix}}_{\text{T}} + {P}^c_{\text{T}}  + P^{\text{deco}}_{cpu}wB\sum_{t=1}^T\log_2(1+\Gamma_t)},\\ 
\text{subject to:} \quad 
& d_{mt} \in \{0,1\}, \quad \forall m \in \mathcal{M}, \ t \in \mathcal{T}, \label{eq_4_1}\\
& s_{m} \in \{0,1\}, \quad \forall m \in \mathcal{M},  \label{eq_4_11}\\
& d_{mt} \leq s_{m}, \quad \forall m \in \mathcal{M}, \ t \in \mathcal{T},  \label{eq_4_12}\\
& \sum\nolimits_{t=1}^{T} d_{mt}  \geq s_{m}, \quad \forall m \in \mathcal{M},  \label{eq_4_13}\\
& 0 \leq \eta^{u}_t \leq 1, \quad \forall t \in \mathcal{T}, \label{eq_4_2}\\
& w\sum\nolimits_{t=1}^{T}\log_2(1+\Gamma_t) \geq \text{SE}^{\text{QoS}}, \label{eq_4_3}\\
& w\log_2(1+\Gamma_t) \geq \text{SE}^{\text{pQoS}}, \quad \forall t \in \mathcal{T}, \label{eq_4_31}\\
& \sum\nolimits_{m=1}^{M} d_{mt} \geq 1, \quad \forall t \in \mathcal{T}, \label{eq_4_4}\\
& \sum\nolimits_{t=1}^{T} d_{mt} \leq T_{\text{max}}, \quad \forall m \in \mathcal{M}, \label{eq_4_41}
\end{align}
\end{subequations}

where $\bm{\eta^{u}} = \{\eta^{u}_t\}_{t {\in} \mathcal{T}}$ is the vector of uplink power control coefficients, $\textbf{D}$ is the AP-UE association matrix with binary elements $d_{mt}$ indicating whether AP $m$ serves UE $t$, and $\textbf{s} {=} \{s_m\}_{m {\in} \mathcal{M}}$ is the AP switching vector where $s_m {=} 1$ indicates that AP $m$ is active.  Constraints \eqref{eq_4_1} and \eqref{eq_4_11} enforce binary decisions for association and activation variables. Constraints \eqref{eq_4_12} and \eqref{eq_4_13} ensure logical coupling between AP activation and association: if an AP is switched off ($s_m {=} 0$), it cannot serve any UEs ($d_{mt} {=} 0$ for all $t$). Constraint \eqref{eq_4_2} ensures that the power control coefficients lie within a feasible range. The term $\text{SE}^{\text{QoS}}$ denotes the minimum total SE required for QoS, while $\text{SE}^{\text{pQoS}}$ represents the per-user minimum SE requirement for fairness. Constraint \eqref{eq_4_3} ensures that the aggregated SE meets the QoS threshold. Constraint \eqref{eq_4_31} ensures per-user fairness by maintaining minimum SE for each UE. Constraint \eqref{eq_4_4} guarantees that each UE is served by at least one AP and Constraint \eqref{eq_4_41} implements fronthaul load control by limiting the maximum number of UEs each AP can serve to $T_{\text{max}}$.

The problem defined in \eqref{eq_4_m} is a mixed-integer nonlinear programming (MINLP) problem. Such problems are generally NP-hard due to their combinatorial and non-convex nature.

\section{Fractional Programming-Based Energy Efficiency Maximization}
\label{solution_analysis}
To address the mixed-integer and non-convex nature of the EE maximization problem in (15), we adopt a fractional programming–based solution framework that combines quadratic transformations with alternating optimization. The proposed approach systematically converts the original problem into a sequence of convex subproblems that can be efficiently solved, while preserving feasibility with respect to the original constraints. The approach is composed of several key steps, as outlined below:
\begin{enumerate}
    \item \textbf{Reformulation of Objective Function:} To reduce the computational cost while preserving optimality, we first reformulate the objective function in \eqref{eq_4_m} to a more tractable form. 
    \item \textbf{Binary Relaxation:} We relax the binary variables to continuous ones, which expands the feasible set and provides an upper bound on the optimal value EE of the original MINLP.  
    \item \textbf{Quadratic Transformation for SINR:} The non-linear SINR expression, \( \Gamma_t \), is made tractable by applying the quadratic transformation method from \cite{shen2018fractional} to obtain the lower bound of \( \Gamma_t \). 
    \item \textbf{Handling Fractional Objective Function:} The fractional structure of the objective functiom, i.e. EE,  is transformed into a tractable form using fractional programming method with the quadratic transformation, as in \cite{shen2018fractional}, to obtain the lower bound on EE.
    \item \textbf{Alternating Optimization and Iterative Refinement:} 
    The coupling between the uplink power-control variables $\bm{\eta^{u}}$ and the AP-UE association and activation variables is addressed via alternating optimization, wherein the resulting convex subproblems are solved iteratively with closed-form updates of the auxiliary variables until convergence.
\end{enumerate}
This structured approach significantly reduces the complexity of solving the EE maximization problem \eqref{eq_4_m} while preserving accuracy and feasibility with respect to the original constraints.

\textbf{Step 1: Reformulating the Objective Function:}
Since the sum SE appears in both the numerator and denominator of \eqref{eq_4}, we reformulate the objective into the equivalent form in \eqref{eq_5_m}, which simplifies the optimization without affecting optimality. A formal proof of equivalence is provided in Appendix \ref{apx_2}.
\begin{subequations}
\label{eq_5_m}
\begin{align} 
\label{eq_5}
\underset{\bm{\eta^{u}},\textbf{D},\textbf{s}}{\max~} &  \frac{wB\sum_{t=1}^{T}\log_2(1+\Gamma_t)}{P^{\text{fix}}_{\text{T}} + {P}^c_{\text{T}}},\\ 
\text{subject to:} \quad
& \Cref{eq_4_1,eq_4_11,eq_4_12,eq_4_13,eq_4_2,eq_4_3,eq_4_31,eq_4_4,eq_4_41}. \label{eq_5_1}
\end{align}
\end{subequations}

\textbf{Step 2: Relaxation of the Binary Association Variable:}
To mitigate the high computational complexity associated with solving MINLP problems, we relax the binary constraints on the variables $d_{mt}$ and $s_{m}$. Specifically, the binary constraints in \eqref{eq_4_1} and \eqref{eq_4_11} is relaxed to a continuous one as follows:
\begin{align}
\label{eq_6}
&0 \leq d_{mt} \leq 1, \quad \forall m \in \mathcal{M}, \  t \in \mathcal{T} \\
& 0 \leq s_{m} \leq 1, \quad \forall m \in \mathcal{M}.\label{eq_6_11}
\end{align}

\textit{Remark 1: Although the binary variables are relaxed to continuous values, the power-consumption structure of the objective inherently penalizes fractional associations, driving the solutions toward near-binary values. This behavior is verified numerically in Section \ref{simulations}.}

\textbf{Step 3: Approximation of the SINR Expression via Auxiliary Variables and Quadratic Transformation:}
The SINR expression \( \Gamma_t \) in both the objective function and the QoS constraints is non-convex due to its fractional structure inside the logarithm. To enable tractable optimization, we introduce an auxiliary variable \( \Gamma^{*}_{t} \) that serves as a concave lower bound to the true SINR and enforce the constraint
\begin{align}
\label{eq_6_1}
\Gamma^{*}_{t} \leq \Gamma_{t}, \quad \forall t \in \mathcal{T}.
\end{align}
However, constraint \eqref{eq_6_1} remains non-convex due to the fractional structure of \( \Gamma_t \). To address this, we apply the quadratic transformation technique on \( \Gamma_t \), which yields the following tractable approximation:
\begin{align}
\label{eq_11}
\Gamma^{*}_{t} \leq 2z_t\sqrt{\mathrm{DS}_t} - z_t^{2}I_t.
\end{align}
where \( I_t = \mathrm{PC}_t + \mathrm{BU}_t + \mathrm{NI}_t + \mathrm{N}_t \) represents the aggregate interference and noise affecting UE \( t \). The quadratic form in \eqref{eq_11} provides a concave lower bound on the original SINR expression \( \Gamma_t \) and becomes tight when the auxiliary variable is updated as
\begin{align}
\label{eq_10}
 z^*_t = \frac{\sqrt{\mathrm{DS}_t}}{I_t}.
\end{align}
This transformation in \eqref{eq_11} preserves feasibility with respect to the original SINR constraint in \eqref{eq_6_1} while enabling efficient optimization within a convex framework.

\begin{pavikl} 
\label{lem_1}
\textsc{:} The quadratic expression in  \eqref{eq_11} is a concave lower bound on the SINR \( \Gamma_t \), and the bound is tight when $z = z^*_t$.

Proof: See Appendix \ref{apx_lemma1}. 
\end{pavikl}

Incorporating the auxiliary variable \( \Gamma^{*}_t \) and its corresponding constraint \eqref{eq_11}, the optimization problem in \eqref{eq_5_m} is reformulated as:
\begin{subequations}
\label{eq_8_m}
\begin{align} 
\label{eq_8}
\underset{\bm{\eta^{u}},\textbf{D},\textbf{s},\bm{z},\bm{\Gamma^{*}}}{\max~} &  \frac{wB\sum\nolimits_{t=1}^{T}\log_2(1+\Gamma^{*}_t)}{P^{\text{fix}}_{\text{T}} + {P}^c_{\text{T}}},\\
\text{subject to:} \quad 
& w\sum\nolimits_{t=1}^{T}\log_2(1+\Gamma^{*}_{t}) \geq \text{SE}^{\text{QoS}}, \quad \forall  t \in \mathcal{T}, \label{eq_8_1} \\
& w\log_2(1+\Gamma^{*}_t) \geq \text{SE}^{\text{pQoS}}, \quad \forall t \in \mathcal{T}, \label{eq_8_11}\\
& \Cref{eq_4_12,eq_4_13,eq_4_2}, \eqref{eq_4_4}, \eqref{eq_4_41}, \eqref{eq_6}, \eqref{eq_6_11}, \eqref{eq_11}. \label{eq_8_3}
\end{align}
\end{subequations}
where \( \bm{z} = \{z_t\}_{t \in \mathcal{T}} \) and \( \bm{\Gamma^{*}} = \{\Gamma^{*}_t\}_{t \in \mathcal{T}} \).
According to Lemma~\ref{lem_1}, as the iterative process refines \( z_t \), the auxiliary variable \( \Gamma^{*}_t \) asymptotically converges to the true SINR \( \Gamma_t \), thus ensuring that Problem \eqref{eq_8_m} yields a solution equivalent to that of the original problem in \eqref{eq_5_m}.

\textbf{Step 4: Handling Fractional Objective Function:}  
The objective function in \eqref{eq_8} has a fractional form, which is non-convex and difficult to optimize directly. To obtain a tractable formulation, we apply a quadratic transformation that converts the energy-efficiency ratio into a concave lower bound. Specifically, we introduce auxiliary variables $u$ and $v$ to represent the numerator and denominator of the EE objective, respectively, and enforce them through the constraints
\begin{align} 
& wB\sum\nolimits_{t=1}^{T}\log_2(1+\Gamma^{*}_t) \geq u, \label{eq_7_5} \\
& P^{\text{fix}}_{\text{T}} + \bar{P}_{\text{T}} \leq v. \label{eq_7_6}
\end{align}
With these inequalities, maximizing EE is equivalent to maximizing the ratio $\frac{u}{v}$. However, the objective function ratio $\frac{u}{v}$ still retains a fractional form. To handle this, we apply the quadratic transformation technique  to derive a concave lower bound
\begin{align}
\label{eq_12}
\frac{u}{v} \geq 2b\sqrt{u} - b^2v,
\end{align}
where \( b \) is an auxiliary variable. This bound becomes tight when
\begin{align}
\label{eq_13}
b^* = \frac{\sqrt{u}}{v}.
\end{align}

\begin{pavikl}
\label{lem_2}
\textsc{:} The quadratic transformation in \eqref{eq_12} provides a concave lower bound on the ratio $\frac{u}{v}$, and the bound is tight when the auxiliary variable $b$ is updated as \( b = b^* \).

\textbf{Proof:} See Appendix \ref{apx_lemma2}.
\end{pavikl}

To enhance numerical stability and avoid infeasibility issues during the iterative solution process, particularly when the problem is later decomposed, we reformulate the hard constraints \eqref{eq_8_1} and \eqref{eq_8_11} as penalty terms in the objective function. By substituting the objective with the lower bound from \eqref{eq_12} and the penalty function, we arrive at the following reformulated problem:
\begin{subequations}
\label{eq_14_m}
\begin{align} 
\label{eq_14}
f =\underset{\bm{\eta^{u}},\textbf{D},\textbf{s},u,v,\bm{z},\bm{\Gamma^{*}}}{\max~} & \ \ 2b\sqrt{u} - b^{2}v - g(\bm{\Gamma^{*}}), \\ 
\text{subject to:} \quad 
& \Cref{eq_4_12,eq_4_13,eq_4_2}, \eqref{eq_4_4}, \eqref{eq_4_41}, \eqref{eq_6},  \eqref{eq_6_11},
 \eqref{eq_11}, \eqref{eq_7_5}, \eqref{eq_7_6}, \label{eq_14_5}
\end{align}
\end{subequations}
where the penalty function $g(\bm{\Gamma^{*}})$ is defined as
\begin{align}
\label{eq_pen}
\begin{split}
 g(\bm{\Gamma^{*}}) & = \lambda_1\max\Big(0, \text{SE}^{\text{QoS}} -w\sum_{t=1}^{T}\log_2(1+\Gamma^{*}_t)   \Big)  + \lambda_2\sum_{t=1}^T\max\Big(0, \text{SE}^{\text{pQoS}} -w\log_2(1+\Gamma^{*}_t)   \Big),
\end{split}
\end{align}
where $\lambda_1$ and $\lambda_2$ are positive penalty constants chosen sufficiently large. The penalty function $ g(\bm{\Gamma^{*}})$ is directly constructed from the original sum and per-user QoS constraints. 
As the iterations proceed, due to large values of $\lambda_1$ and $\lambda_2$, the penalty term  $g(\bm{\Gamma^{*}})$ approaches near zero value, satisfying the original constraints \eqref{eq_8_1} and \eqref{eq_8_11}. Simultaneously, the value of $2b\sqrt{u} - b^{2}v$  approaches that of the fractional form $\frac{u}{v}$, as established in Lemma \ref{lem_2}. This implies that solving Problem \eqref{eq_14_m} is equivalent to solving Problem \eqref{eq_8_m}. The resulting penalized objective is then optimized using alternating optimization over the power control and association variables, as detailed in the next step.

\textbf{Step 5: Alternating Optimization and Iterative Refinement:}
After the quadratic transformations in Steps~3 and~4, the remaining non-convexity arises from the coupling between the uplink power-control variables and the AP-UE association and activation variables. This coupling is addressed using an alternating optimization strategy.

For fixed AP--UE association and activation variables $\{\mathbf{D},\mathbf{s}\}$, the resulting subproblem in~\eqref{eq_15_m} is convex with respect to the uplink power-control variables $\bm{\eta}^{u}$ and the auxiliary variables, and can be efficiently solved to update $\bm{\eta}^{u}$. Subsequently, for fixed uplink power-control variables, the association and activation subproblem in~\eqref{eq_16_m} becomes convex and is solved to update $\mathbf{D}$ and $\mathbf{s}$. The auxiliary variables are updated in closed form between successive iterations. Specifically, the two convex subproblems solved at each iteration are given by
\begin{subequations}
\label{eq_15_m}
\begin{align} 
\label{eq_15}
f_1 = \underset{\bm{\eta^{u}},u,v,\bm{\Gamma^{*}}}{\max~} & \ \ 2b\sqrt{u} - b^{2}v - g(\bm{\Gamma^{*}}), \\ 
\text{subject to:} \quad 
&  \eqref{eq_4_2},  \eqref{eq_11}, \eqref{eq_7_5}, \eqref{eq_7_6}. \label{eq_15_1}
\end{align}
\end{subequations}
\begin{subequations}
\label{eq_16_m}
\begin{align} 
\label{eq_16}
f_2 = \underset{\textbf{D},\textbf{s},u,v,\bm{\Gamma^{*}}}{\max~} & \ \ 2b\sqrt{u} - b^{2}v - g(\bm{\Gamma^{*}}), \\ 
\text{subject to:} \quad 
& \eqref{eq_4_12}, \eqref{eq_4_13},  \eqref{eq_4_4}, \eqref{eq_4_41}, \eqref{eq_6}, \eqref{eq_6_11}, \eqref{eq_11}, \eqref{eq_7_5}, \eqref{eq_7_6}. \label{eq_16_1}
\end{align}
\end{subequations}
This alternating procedure between the two sub-problems continues until convergence. The complete iterative procedure is summarized in Algorithm~\ref{Algorithm 1}.

Collectively, the solution approach constructs a system of progressively tightened bounds. Step 1 maintains the optimality, as shown in Appendix \ref{apx_2}. In Step 2, the binary relaxed NLP problem provides an upper bound on the original MINLP's optimal EE. This upper bound is tight, as evidenced by the natural binarization of the relaxed variables (Fig. \ref{fig:relaxation_validation}) and the negligible performance loss after rounding (Fig. \ref{fig:EE_SE_rounding_off}). Steps 3-4 establish concave lower bounds that become tight in each iteration when auxiliary variables ${z}_t$ and $b$ are set to their optimal values, as proven in Lemmas \ref{lem_1} and \ref{lem_2}. The alternating optimization in Steps 5 then updates the primal variables by solving decoupled convex problems, leading to monotonic improvement and convergence. 

\begin{algorithm}[]
\caption{Proposed Alternating Optimization Algorithm}\label{Algorithm 1}
\begin{algorithmic}[1]
\STATE \textbf{Initialization:} Initialize feasible values ${\bm{\eta^{u}}}^{(1)}$, $\textbf{D}^{(1)}$, $\textbf{s}^{(1)}$ and set iteration index \( i = 1 \), tolerance \( \epsilon = 5 \times 10^{-3} \).
\REPEAT
    \FORALL{ \( t \in \mathcal{T} \) }
        \STATE Calculate $\textbf{a}_t$ using \eqref{eq:LSFD}.
        \STATE Update \( z_t \) using \eqref{eq_10}.
    \ENDFOR
    \STATE Update \( b \) using \eqref{eq_13}.
    \STATE Solve Problem \eqref{eq_15_m} to update \( {\bm{\eta^{u}}}^{(i+1)} \).
    \FORALL{ \( t \in \mathcal{T} \) }
        \STATE Recompute \( z_t \) using \eqref{eq_10}.
    \ENDFOR
    \STATE Recompute \( b \) using \eqref{eq_13}.
    \STATE Solve Problem \eqref{eq_16_m} to update \( \textbf{D}^{(i+1)} \) and \( \textbf{s}^{(i+1)} \).
    \STATE \( i \gets i + 1 \).
\UNTIL{ \( \left|\frac{f^{(i)} - f^{(i-1)}}{f^{(i-1)}}\right| \leq \epsilon \). }
\end{algorithmic}
\end{algorithm}

\textit{Complexity Analysis}: The computational cost of each iteration of Algorithm~\ref{Algorithm 1} is dominated by solving the two convex subproblems \eqref{eq_15_m} and \eqref{eq_16_m}. Using the interior-point method for convex optimization, the cost of solving a convex problem is $\mathcal{O}(\sqrt{r_l + r_q}(r_v + r_l + r_q)r_v^2)$ \cite{hao2024joint,tam2016joint}, where $r_v$ is the number of scalar variables, $r_l$ is the number of linear constraints, and $r_q$ is the number of quadratic constraints. Thus, the cost of solving \eqref{eq_15_m} is $\mathcal{O}(4\sqrt{4T + 1}(6T+5)(T+1)^2) \approx \mathcal{O}(T^{3.5})$ and  the cost of solving \eqref{eq_16_m} is $\mathcal{O}(\sqrt{3MT+4M+5T+3}(4MT+5M+6T+5)(MT+M+T+2)^2) \approx  \mathcal{O}(M^{3.5}T^{3.5})$. 
For the numerical results, convex subproblems are solved using an interior-point method for performance evaluation only. Since all subproblems are convex after the proposed transformations, they can be efficiently solved using first-order methods in practical implementations. These methods have significantly lower per-iteration complexity and are well suited for large-scale D-mMIMO networks.

\textit{Convergence Analysis}:  The convergence of the proposed alternating optimization in Algorithm~\ref{Algorithm 1} to a stationary point of the original problem \eqref{eq_4_m} is established under the following assumptions:
\begin{itemize}
    \item The feasibility of the original problem \eqref{eq_4_m} is preserved at each iteration of Algorithm~\ref{Algorithm 1}.
    \item The objective function exhibits monotonic improvement at every iteration.
    \item The objective functions of sub-problems \eqref{eq_15_m} and \eqref{eq_16_m} are bounded from above.
\end{itemize}

\begin{pavikt}
\label{t_1}
\textsc{:} Let the sequence $\{{\bm{\eta^u}}^{(i)}, \textbf{D}^{(i)}, \textbf{s}^{(i)}\}$ be generated by Algorithm~\ref{Algorithm 1}. Then, under the above assumptions 1--3 the sequence $\{{\bm{\eta^u}}^{(i)}, \textbf{D}^{(i)}, \textbf{s}^{(i)}\}$ converges to a limit point $\{{\bm{\eta^u}}^{*}, \textbf{D}^{*}, \textbf{s}^{(*)}\}$, which is a stationary point of the original problem~\eqref{eq_4_m}.

Proof: See Appendix~\ref{apx_3}.
\end{pavikt}

\section{Numerical Simulations}
\label{simulations}
In this section, numerical results are presented to evaluate the proposed EE maximization framework. Rather than focusing solely on numerical performance, the discussion emphasizes system-level insights,
highlighting when joint AP activation, AP-UE association, and power control significantly affect the EE-SE trade-off in uplink D-mMIMO systems.

We conduct numerical simulations over a geographical area where APs and UEs are independently and identically distributed over a square $1 \times 1$~km area, centred at the origin. Specifically, the $x$ and $y$ coordinates of each UE and AP are independently drawn from a continuous uniform distribution ranging from -0.5 to 0.5.
To emulate an infinitely large network and eliminate edge effects, we adopt a wrap-around topology as described in \cite{bjornson2020scalable}. In our setup, the number of APs is denoted by $M$, and the number of UEs is $T$.

\begin{table*}[htbp]
\centering
\footnotesize
\caption{System Parameters}
\label{tab:system_parameters}
\renewcommand{\arraystretch}{1}
\begin{tabular}{|c|c|c|c|c|c|c|c|}
\hline
\textbf{Symbol} & \textbf{Value} & \textbf{Symbol} & \textbf{Value} & \textbf{Symbol} & \textbf{Value}& \textbf{Symbol} & \textbf{Value} \\
\hline
$B$ & $20$ MHz & $A$ & $8$ & $T_{\text{max}}$ & $10$ &$L_c$, $L_p$ &$200$,  $5$\\
  $p_p$ & $100$ mW & $p_u$ & $100$ mW  &$\text{SE}^{\text{pQoS}}$ & $0.1$ Bits/s/Hz & $\text{SE}^{\text{QoS}}$ & $100$ Bits/s/Hz \\
$\zeta$ & $0.4$ & $P^{\text{c}}_{ue}$ & $100$ mW & $P^{\text{c}}_{ap}$ & $100$ mW & $P^{\text{proc}}$ & $800$ mW \\
$P^{\text{fix}}_{fh}$ & $825$ mW & $P^{\text{sig}}$ & $10$ mW & $P^{\text{fix}}_{cpu}$ & $5000$ mW &
$P^{\text{lsfd}}_{cpu}$ & $1000$ mW \\
 $P^{\text{deco}}_{cpu}$ & $1000$ mW/Gb/s & Shadow fading & $8$ dB  
& $\lambda_1$ & $1 \times 10^4$ & $\lambda_2$ & $ 2 \times 10^4$
\\
\hline
\end{tabular}
\end{table*}

The power-consumption parameters used in Table \ref{tab:system_parameters} are adopted from \cite{chen2023energy}.
The LSFCs are modeled using a three-slope path loss model, with shadow fading applied using a log-normal distribution~\cite{ngo2017cell}. Unless specified otherwise, we fix the system parameters as given in Table \ref{tab:system_parameters}. Pilots are assigned using the error estimation minimization pilot assignment (EEM-PA) strategy from~\cite{khan2025pilot}.  Unless stated otherwise, all other parameters follow the configuration described in \cite{ngo2017cell}. All numerical results are averaged over 100 independent simulation realizations, ensuring a 95\% confidence interval narrower than ±0.02.

\textit{Note: After convergence of the relaxed optimization problem, the AP–UE association and AP activation variables are obtained by rounding-off with a threshold of $0.5$. The resulting discrete solution is then checked against the original QoS constraints. We employ a conservative performance evaluation methodology: any realization that fails to satisfy either the per-user SE constraints (fundamental service guarantees) for more than 10\% of users or the sum SE requirement (system-level performance target) is declared infeasible and assigned zero EE. This reflects the fact that systems failing to meet required quality-of-service levels cannot be regarded as energy-efficient, regardless of the resources consumed.}

\subsection{Numerical analysis of the proposed scheme}
\begin{figure}[!h]
    \centering
    \includegraphics[width=0.60\textwidth]{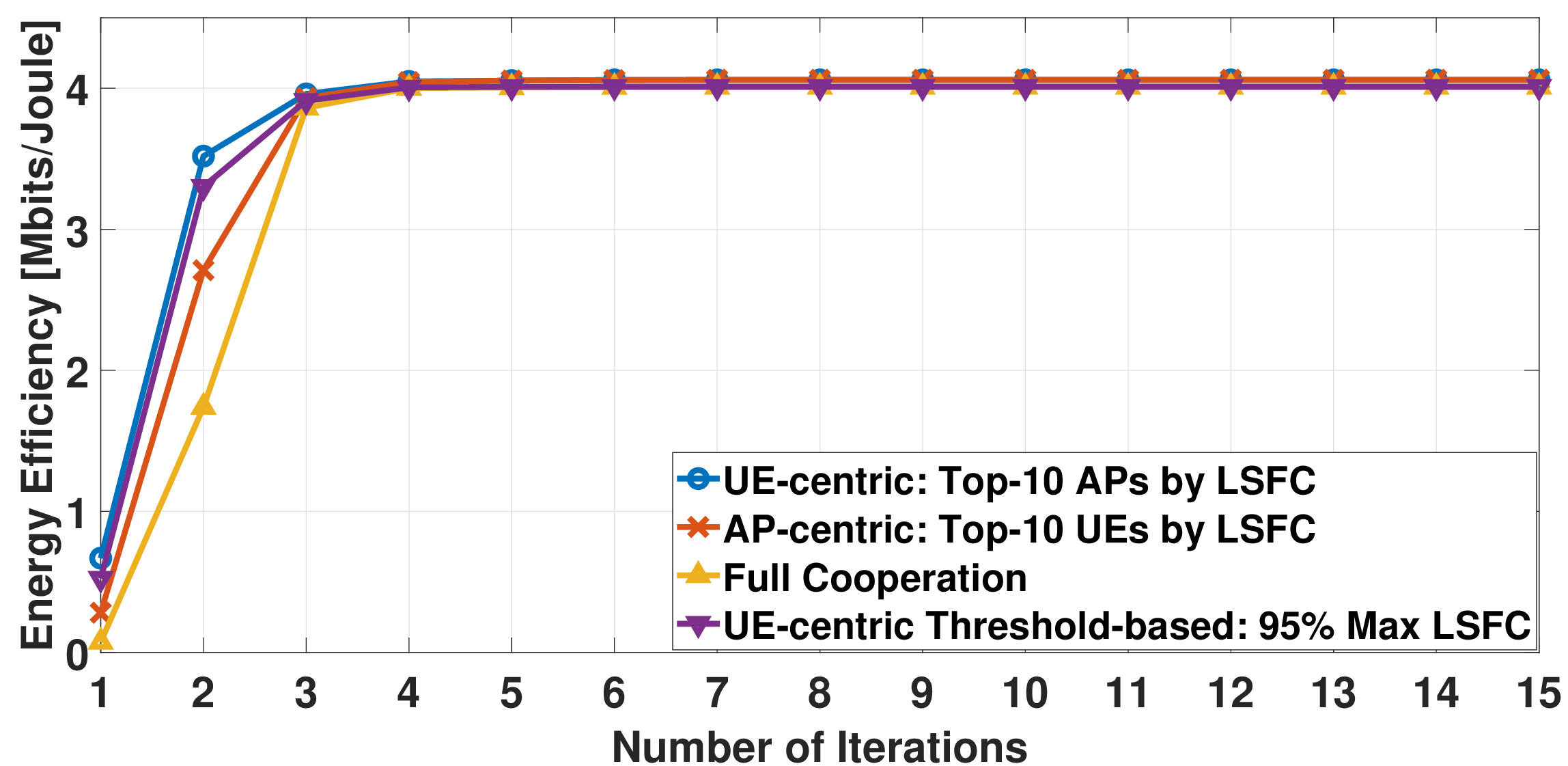}
    \caption{Convergence of EE versus the number of iterations for different AP-UE association initialization strategies for $T=40$ and $M=100$.}
    \label{fig:ap_init}
\end{figure}

\begin{figure}[!h]
    \centering
    \includegraphics[width=0.60\textwidth]{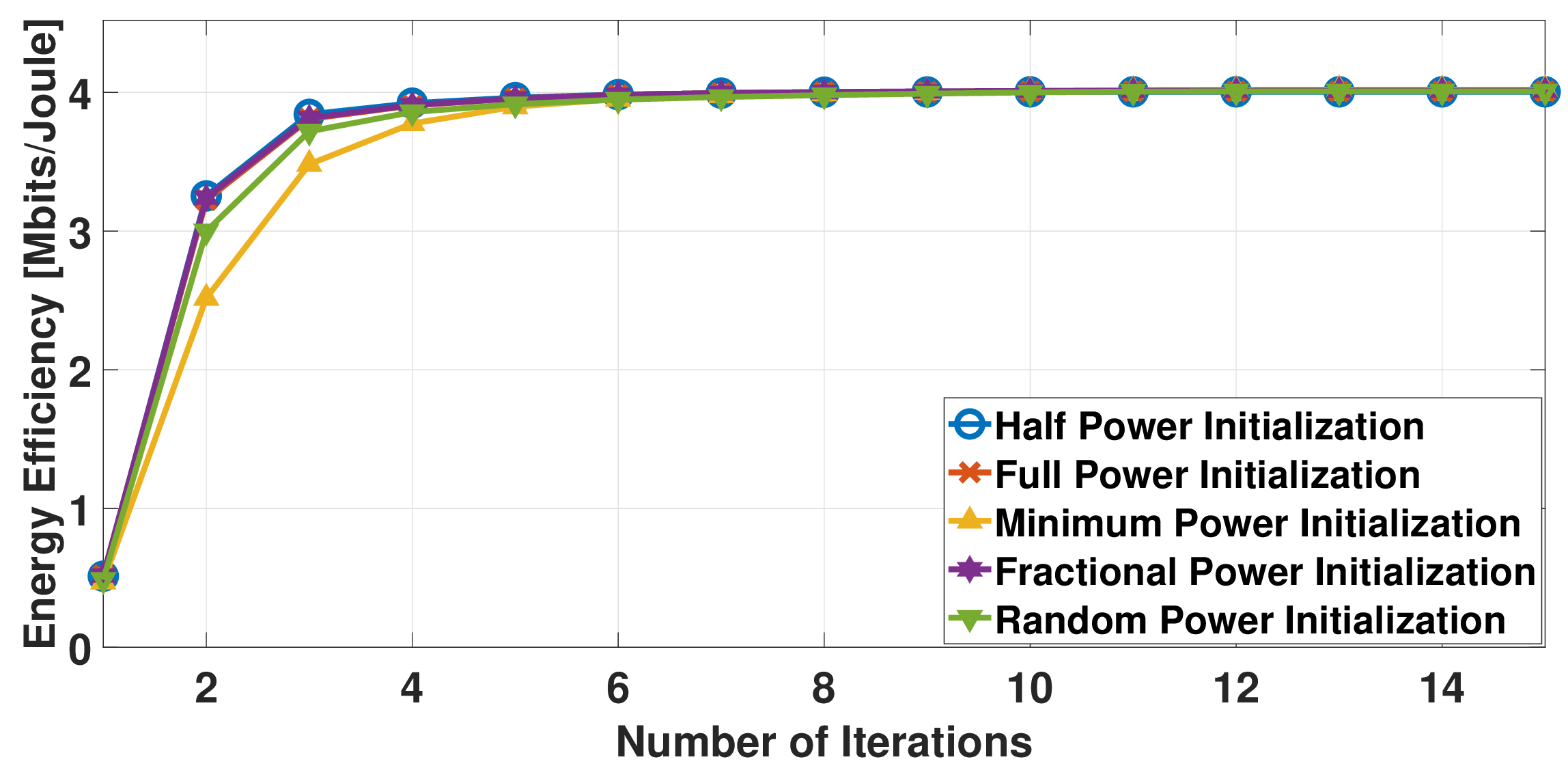}
    \caption{Convergence of EE versus the number of iterations for different power initialization for $T=40$ and $M=100$.}
    \label{fig:power_init}
\end{figure}
The convergence behavior of the proposed iterative algorithm is illustrated in Fig.~\ref{fig:ap_init} and Fig.~\ref{fig:power_init} for different initializations of the AP-UE association and uplink power allocation, respectively. In all cases, the algorithm converges rapidly within 3--6 iterations, indicating low sensitivity to initialization. This fast convergence can be attributed to the structure of the proposed solution framework, where the combination of quadratic transformations and alternating
optimization yields a sequence of well-conditioned convex subproblems. In addition, the presence of a binding sum SE constraint defines a compact feasible region, which guides the optimization trajectory toward the same stationary point regardless of the initial conditions. Notably, the final achieved energy efficiency varies by less than 1\% across all initialization strategies, and the convergence speed differs by at most 1--2 iterations. These results demonstrate that the proposed algorithm is both robust and computationally efficient, making it suitable for practical implementation on network-control timescales.

\begin{figure}[!h]
    \centering
    \includegraphics[width=0.60\textwidth]{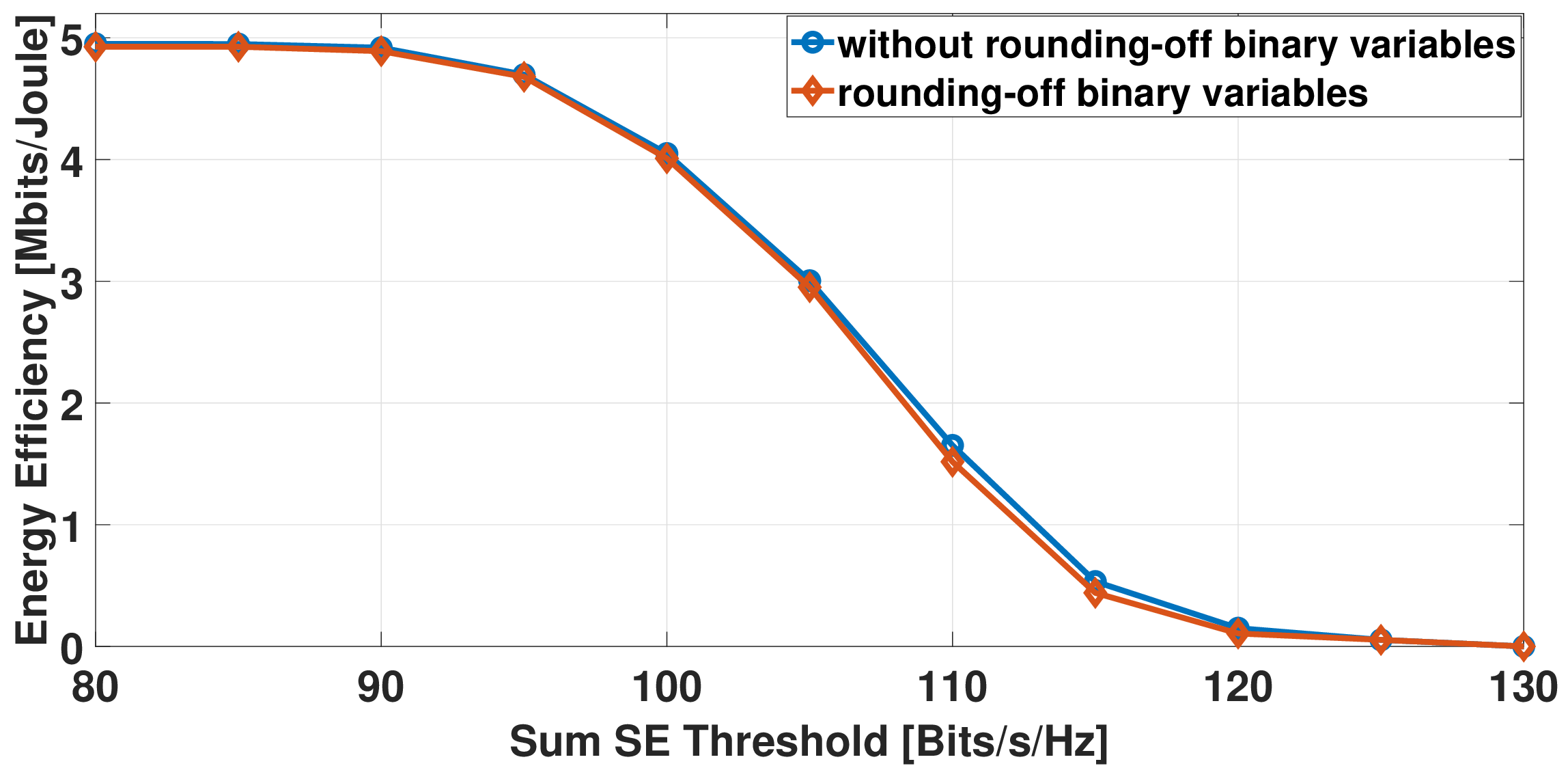}
    \caption{Energy efficiency versus sum SE threshold with and without rounding-off binary variables, when $T=40$ and $M=100$}
    \label{fig:EE_SE_rounding_off}
\end{figure}
Fig.~\ref{fig:EE_SE_rounding_off} illustrates the impact of rounding the relaxed binary variables for AP activation and AP–UE association on the achieved EE for various sum SE thresholds. In our formulation, binary decision variables are relaxed to continuous ones, and the incorporated power-based penalty terms naturally drive these relaxed variables toward near-binary values. The comparison shows that the EE obtained after simple rounding closely matches the EE computed directly from the relaxed solution. This close alignment confirms that the relaxation–rounding strategy provides an accurate and tractable approximation of the original mixed-integer problem without noticeable performance loss, validating both the efficiency and the practicality of the proposed optimization framework.

\begin{figure}[!h]
    \centering
    \includegraphics[width=0.60\textwidth]{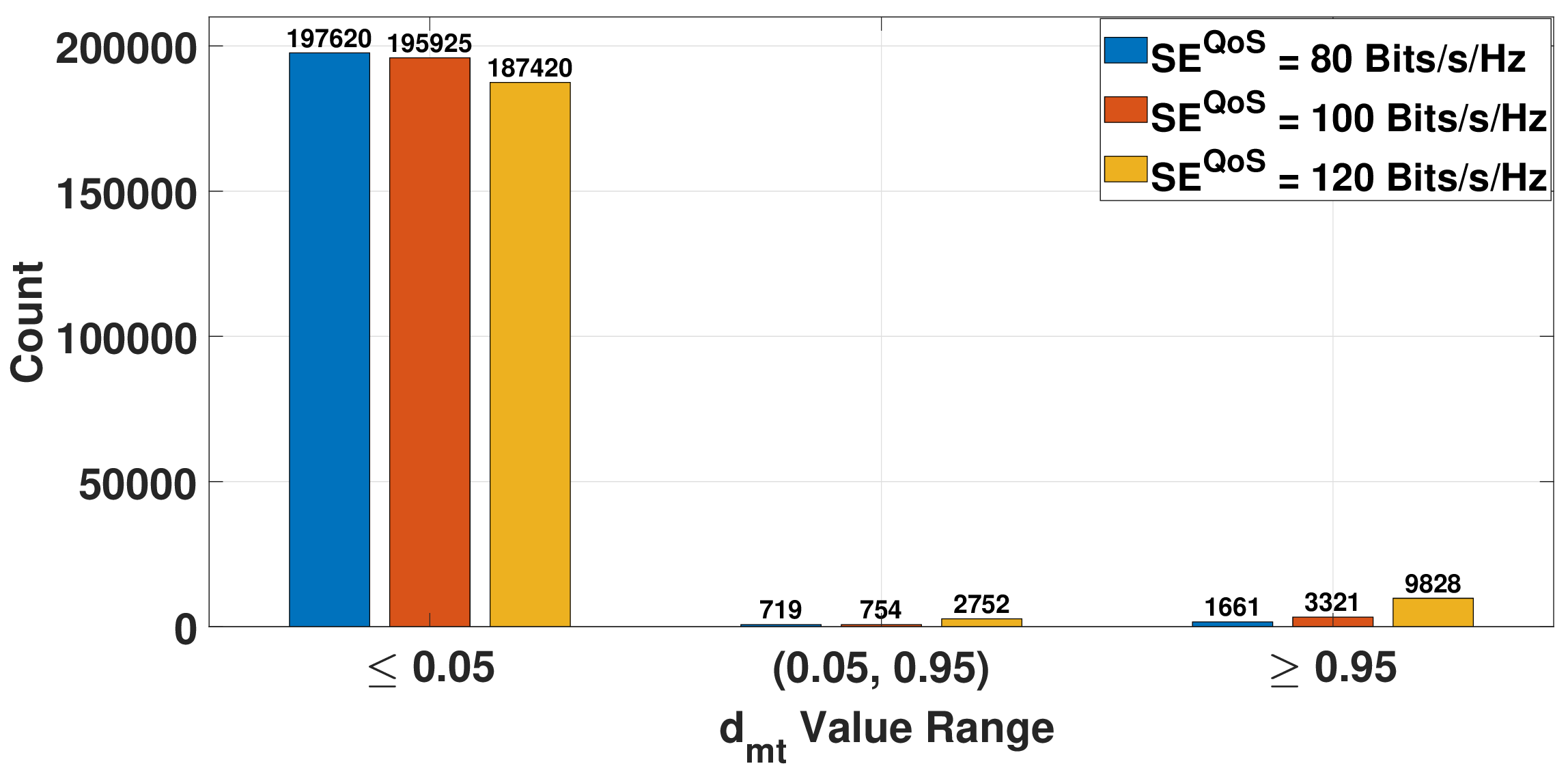}
    \caption{Distribution of optimized AP-UE association variables $d_{mt}$ after convergence for different sum SE thresholds.}
    \label{fig:relaxation_validation}
\end{figure}
Fig.~\ref{fig:relaxation_validation} shows the distribution of optimized AP-UE association variables $d_{mt}$ values across three different sum SE thresholds. For $\text{SE}^{\text{QoS}}$ = 80, 100, and 120 bits/s/Hz, the percentages of $d_{mt}$ variables converging to within 0.05 of binary values (0 or 1) are {99.64\%}, {99.62\%}, and {98.62\%}, respectively. This strong natural binarization occurs because the power consumption terms in the objective function inherently penalize fractional associations, driving variables toward discrete solutions. The observed trend across SE thresholds reveals important system behavior: as throughput requirements increase from 80 to 120 bits/s/Hz, the percentage of active AP-UE connections ($d_{mt} > 0.95$) increases from 0.83\% to 4.91\%, while inactive connections ($d_{mt} \leq 0.05$) decrease from 98.81\% to 93.71\%. This demonstrates that higher SE demands necessitate more extensive AP-UE associations, and our optimization framework adapts accordingly. The binary nature of AP activation variables $s_m$ is inherently influenced by $d_{mt}$ through constraints \eqref{eq_4_12} and \eqref{eq_4_13}, which couple AP activation with AP-UE associations.

\subsection{Performance Evaluation}
For comprehensive performance evaluation, we consider two categories of comparison schemes. First, we include reference schemes that differ fundamentally from the proposed framework either in network architecture or optimization objective. Second, we consider benchmark schemes that share the same uplink D-mMIMO architecture and energy-efficiency objective, but restrict the set of optimized
variables.

\begin{itemize}
\item \textbf{Small-Cell}: A conventional cellular architecture where each UE is served by exactly one AP. The optimization follows~\eqref{eq_4_m} with the additional constraint $\sum_{m=1}^{M} d_{mt}=1, \forall t \in \mathcal{T}$.

\item \textbf{Fixed Association} \cite{vu2020joint,nguyen2020spectral}: Energy efficiency is maximized while fixing the AP-UE association based on LSFCs, where each UE is served by APs contributing at least 95\% of its total received LSFC. Uplink power allocation and AP activation
are optimized.

\item \textbf{Fully Static} \cite{bashar2019energy}:
Energy efficiency is maximized with both AP activation and AP--UE association fixed using the same LSFC-based rule, and only uplink power allocation is optimized under realistic power consumption and sum SE constraints.

\item \textbf{No AP Activation} \cite{khan2024joint}: Energy efficiency is maximized through joint optimization of uplink power allocation and AP-UE association under the same constraints as the proposed framework, while keeping all APs active ($s_m = 1, \forall m$).

\item \textbf{Sum SE Maximization}: The sum spectral efficiency is maximized using the same optimization variables and constraints as the proposed method, with AP-UE association discretization enforced via an $\ell_1$-norm penalty.
\end{itemize}

\begin{figure}[!h]
    \centering
    \includegraphics[width=0.60\textwidth]{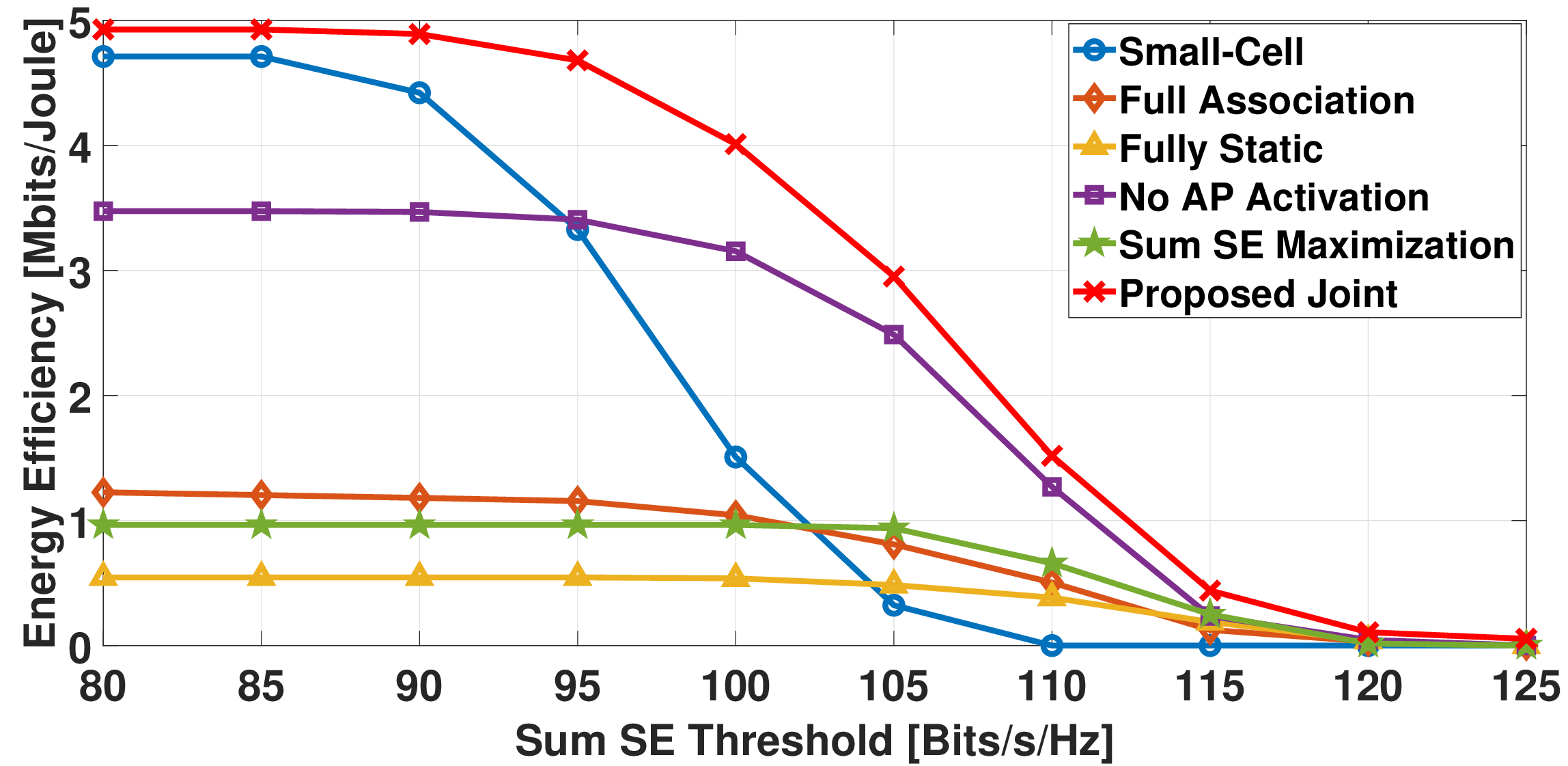}
    \caption{Energy efficiency versus spectral efficiency threshold ($\text{SE}^{\text{QoS}}$) for different benchmarking schemes, when $T=40$ and $M=100$.}
    \label{fig:comparison}
\end{figure}
Fig.~\ref{fig:comparison} illustrates the EE achieved by the considered schemes as a function of the sum SE threshold $\text{SE}^{\text{QoS}}$. For all schemes, the achievable EE decreases as the SE requirement increases, since higher throughput targets necessitate increased transmit power and the activation of additional APs. 

At low SE thresholds, the proposed joint optimization achieves only a modest EE gain over the \textbf{Small-Cell} architecture. In this regime, the throughput requirement can be satisfied with single-AP transmission, leaving limited room for energy savings through multi-AP cooperation. As the SE threshold increases to moderate levels, the EE advantage of the proposed method becomes pronounced. This is because the \textbf{Small-Cell} architecture cannot exploit macro-diversity, whereas the proposed framework selectively activates and associates multiple APs only when doing so is energy efficient. Consequently, the proposed scheme sustains substantially higher EE and supports significantly larger SE targets, while the \textbf{Small-Cell} configuration becomes infeasible at high SE requirements.

Compared with the \textbf{No AP Activation} method, the proposed method achieves higher EE at low-to-moderate SE thresholds due to its ability to deactivate energy-inefficient APs and reduce circuit power consumption. As the SE requirement becomes stringent, nearly all APs must remain active, causing the performance gap between the two schemes to narrow. The proposed framework also consistently outperforms the \textbf{Fixed Association} and \textbf{Fully Static} schemes, highlighting the importance of adaptive AP-UE association and dynamic AP activation for energy-efficient operation. In addition, it achieves higher EE than the Sum SE Maximization scheme, which prioritizes SE without explicitly accounting for infrastructure energy consumption. Overall, the results demonstrate that coordinated AP activation, adaptive association, and power control are essential for balancing EE and SE in uplink D-mMIMO systems across a wide range of throughput requirements.

\begin{figure}[!h]
    \centering
    \includegraphics[width=0.60\textwidth]{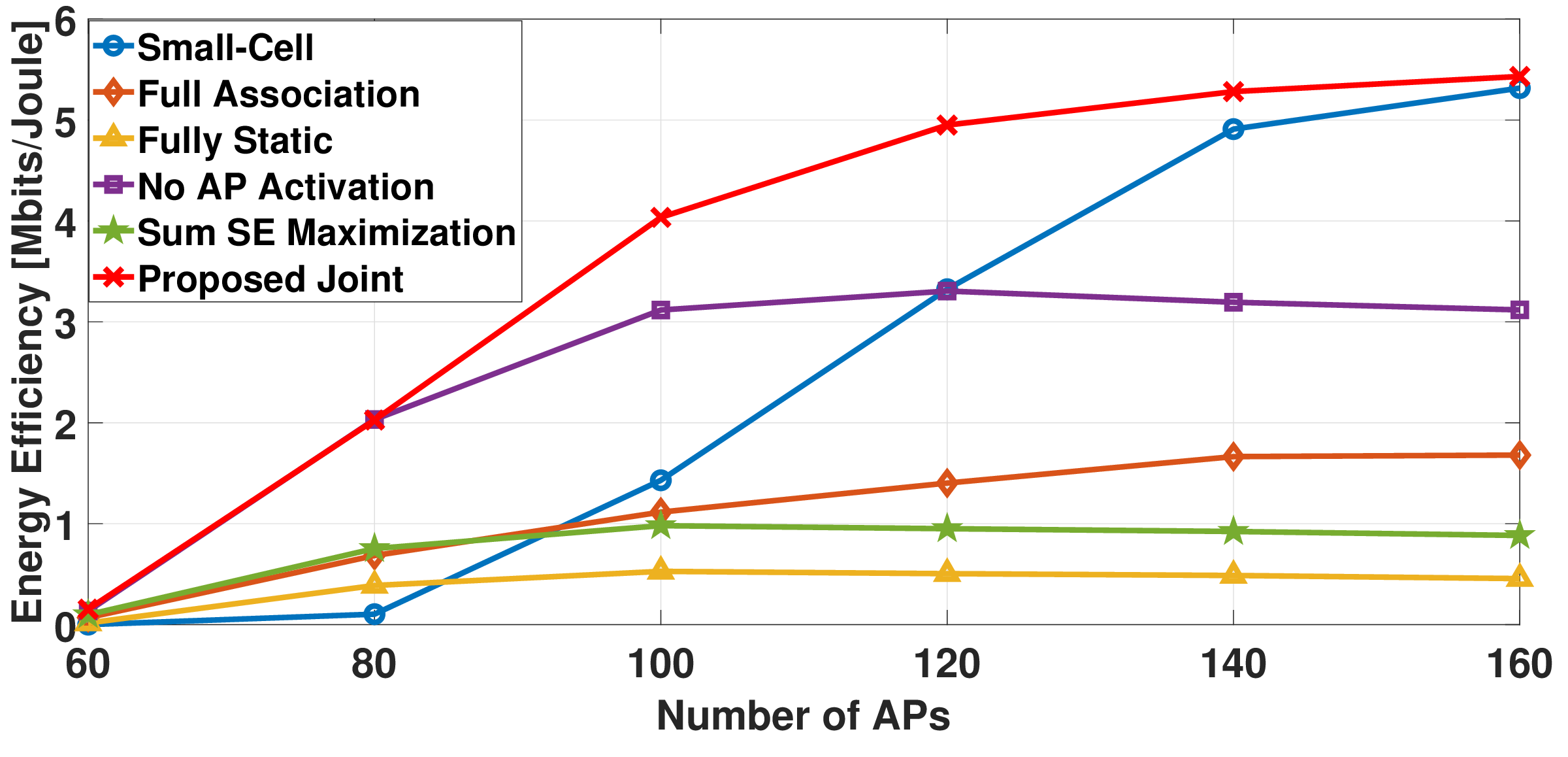}
    \caption{Energy efficiency versus number of APs for different benchmarking schemes, when $T=40$ and $\text{SE}^{\text{QoS}}=100$.}
    \label{fig:comparison_AP}
\end{figure}
Fig.~\ref{fig:comparison_AP} illustrates the EE achieved by the considered schemes as a function of the number of APs under a fixed sum SE requirement of $\text{SE}^{\text{QoS}}=100$~bits/s/Hz. When the number of deployed APs is small, meeting the SE target requires activating most available APs or associating multiple APs per UE, which increases circuit and processing power consumption and limits the achievable EE. In this low-AP-density regime, the proposed method and the \textbf{No AP activation} method exhibit similar EE performance, since AP activation flexibility is inherently limited when nearly all APs must remain active. As the number of APs increases, the EE achieved by the proposed framework improves consistently, owing to its ability to exploit favorable channel conditions while selectively deactivating energy-inefficient APs and adapting the AP-UE associations. In contrast, the EE of \textbf{Fixed Association} and \textbf{No AP Activation} schemes initially benefits from improved channel gains and spatial coverage, but eventually saturates or degrades as infrastructure-related power consumption becomes dominant. The \textbf{Small-Cell} architecture exhibits a gradual EE increase with AP density, since shorter AP-UE distances improve channel quality; however, its lack of multi-AP cooperation limits the achievable gains.

Overall, the results demonstrate that dense AP deployments can significantly improve uplink EE only when combined with adaptive AP activation and AP-UE association. By jointly optimizing these dimensions, the proposed framework consistently achieves the highest EE across all AP densities.

\begin{figure}[!h]
    \centering
    \includegraphics[width=0.60\textwidth]{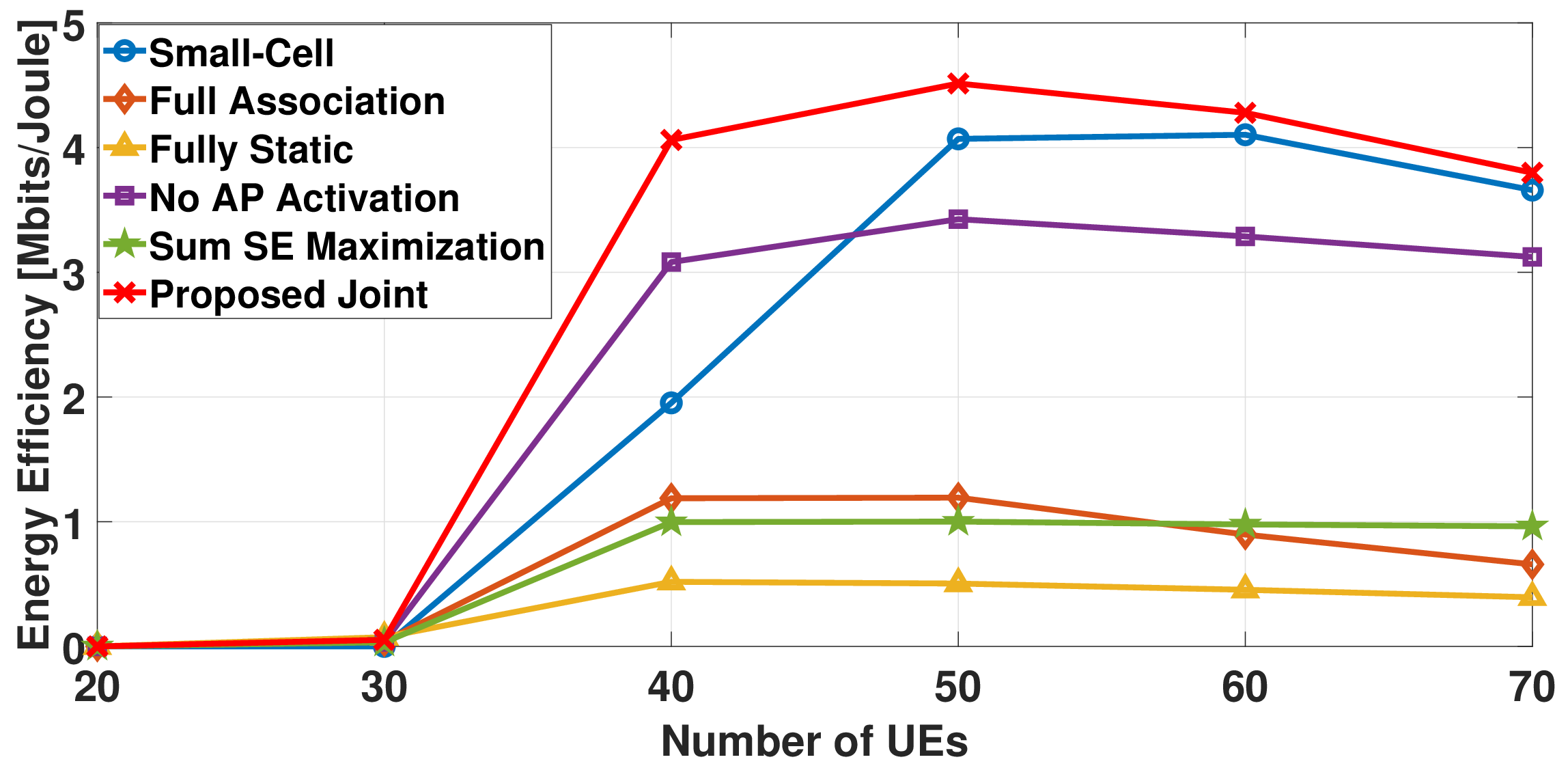}
    \caption{Energy efficiency versus number of UEs for different benchmarking schemes, when $M=100$ and $\text{SE}^{\text{QoS}}=100$.}
    \label{fig:comparison_UE}
\end{figure}
Fig.~\ref{fig:comparison_UE} illustrates the EE achieved by the considered schemes as a function of the number of UEs for $M=100$ and $\text{SE}^{\text{QoS}}=100$. As the number of UEs increases, the sum SE initially improves due to enhanced multi-user diversity, while the associated increase in transmit and processing power remains moderate. In this regime, the EE improves across all schemes. Beyond a moderate UE load, however, the marginal SE gains diminish while the required transmit, decoding, and processing power increase more rapidly, leading to a decline in EE for all schemes. This behavior highlights the fundamental trade-off between throughput gains and infrastructure-related power consumption in uplink D-mMIMO systems. The \textbf{Small-Cell} architecture achieves relatively low EE at small UE counts due to its inability to exploit multi-AP diversity and cooperation. As the number of UEs increases, resource utilization improves and the \textbf{Small-Cell} configuration becomes capable of meeting the sum SE requirement, causing its EE to approach that of the proposed framework. Nevertheless, due to limited spatial diversity and less flexible resource allocation, it consistently attains lower EE than the proposed method.

In contrast, the proposed joint optimization framework maintains superior EE across all user densities by dynamically adapting AP activation and AP-UE association decisions to the offered load. These results demonstrate that load-aware infrastructure adaptation is essential for sustaining energy-efficient uplink operation under varying traffic conditions.

\begin{figure}[!h]
    \centering
    \includegraphics[width=0.60\textwidth]{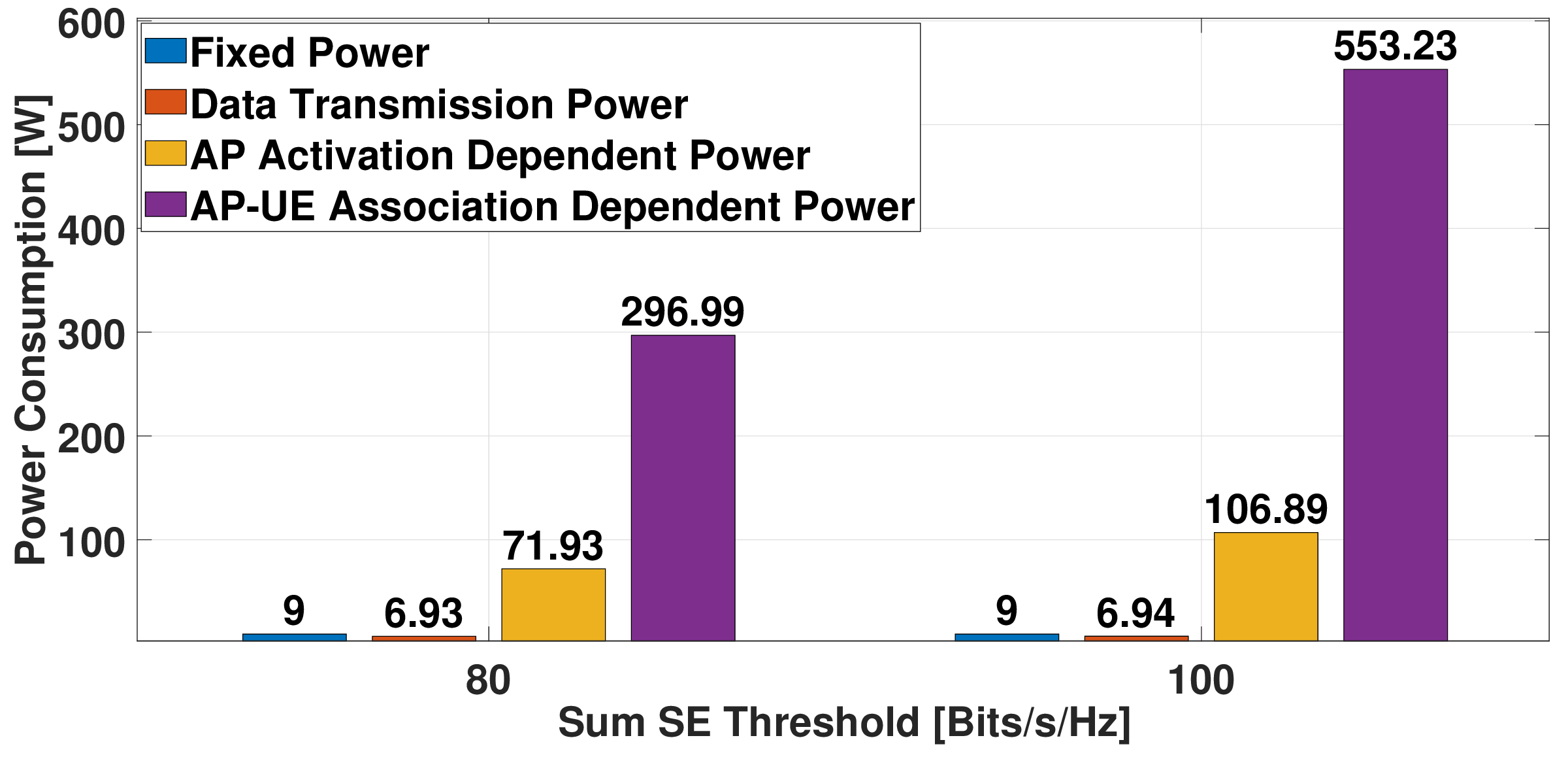}
    \caption{Breakdown of total power consumption under different sum SE thresholds, when $T=40$ and $M=100$}
    \label{fig:power_SE}
\end{figure}
Fig.~\ref{fig:power_SE} presents the breakdown of total power consumption for two different sum SE thresholds (80 and 100~bits/s/Hz). It can be observed that the AP–UE association dependent power dominates the total energy usage, accounting for the majority of power consumption in both cases. As the SE threshold increases from 80 to 100~bits/s/Hz, this component rises sharply from 296.99~W to 553.23~W, indicating the additional fronthaul and signal processing required to support higher SE. The AP activation dependent power also increases moderately (from 71.93~W to 106.89~W) as more APs are activated to meet the higher SE demand. In contrast, the fixed and data transmission power components remain almost constant at around 9~W and 7~W, respectively, and thus have negligible influence on the overall trend. These results clearly show that AP–UE association and activation dynamics play a crucial role in determining the total energy footprint in distributed MIMO systems.

\begin{figure}[!h]
    \centering
    \includegraphics[width=0.60\textwidth]{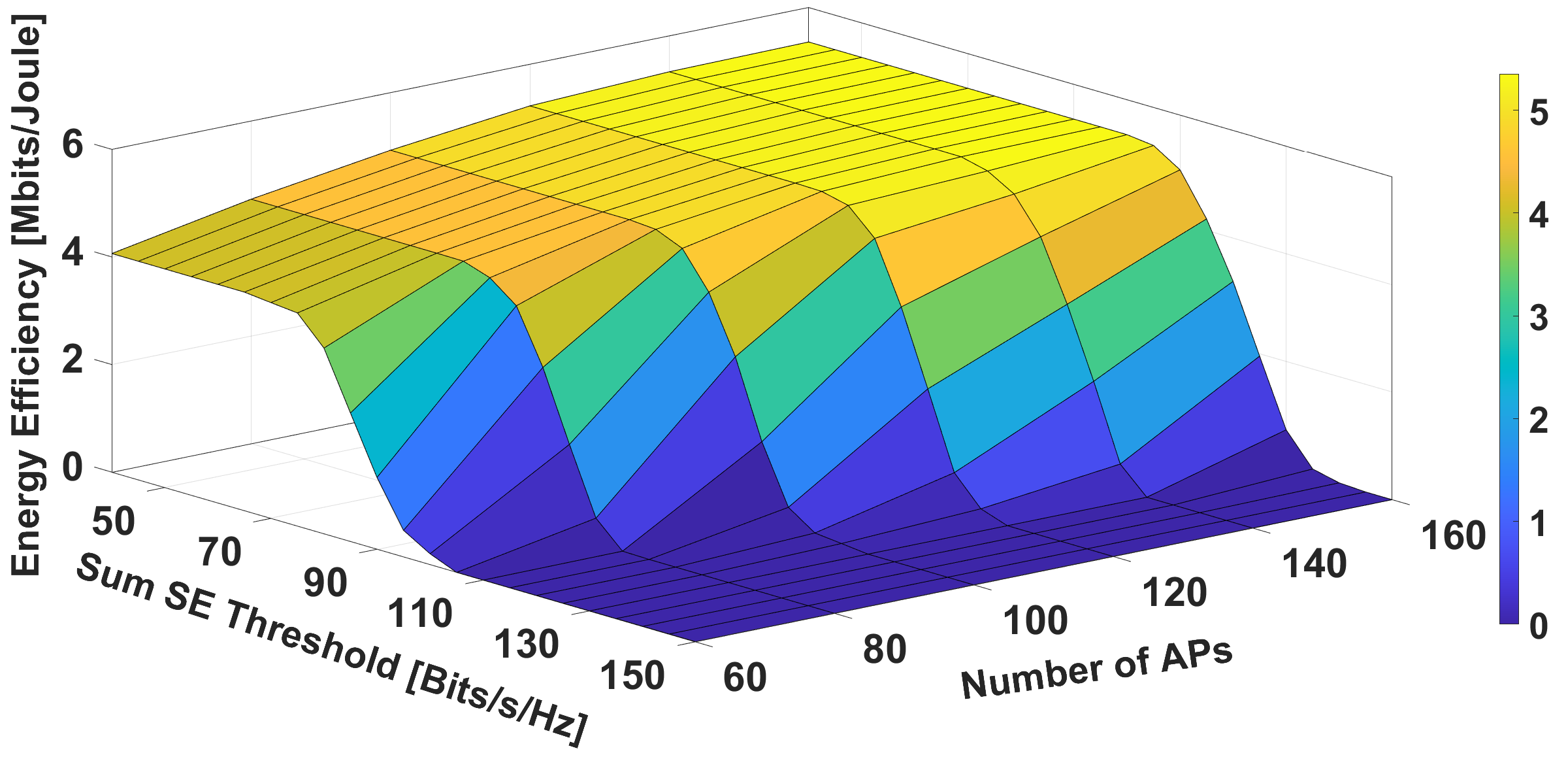}
    \caption{3D surface plot of EE as a function of the number of APs and the sum SE threshold, when $T=40$.}
    \label{fig:EE_SE_AP}
\end{figure}
Fig.~\ref{fig:EE_SE_AP} illustrates the variation of EE with the number of APs and the sum SE threshold. The results highlight a clear EE–SE trade-off and its dependence on AP density. For a fixed number of APs, increasing the sum SE requirement results in a consistent decline in EE. The demand for higher throughput necessitates activating more APs and more aggressive AP-UE association, which raises total energy consumption. Importantly, the systems with a larger number of APs are able to support higher sum SE threshold demands, due to improved spatial diversity. In contrast, network with fewer APs become infeasible at high sum SE thresholds. For sum SE thresholds (up to 100~bits/s/Hz), the system achieves its maximum EE of approximately 5.3~Mbits/Joule at around 120 APs. Beyond this point, further increasing AP density results in EE saturation, as additional APs contribute minimal gain while power consumption increases too with similar rate. For higher SE thresholds, this optimal point shifts to larger AP counts. Overall, these results highlight that the proposed method scales efficiently with AP density, maintaining high EE and enabling support for demanding SE targets through adaptive resource coordination. 

\begin{figure}[!h]
    \centering
    \includegraphics[width=0.60\textwidth]{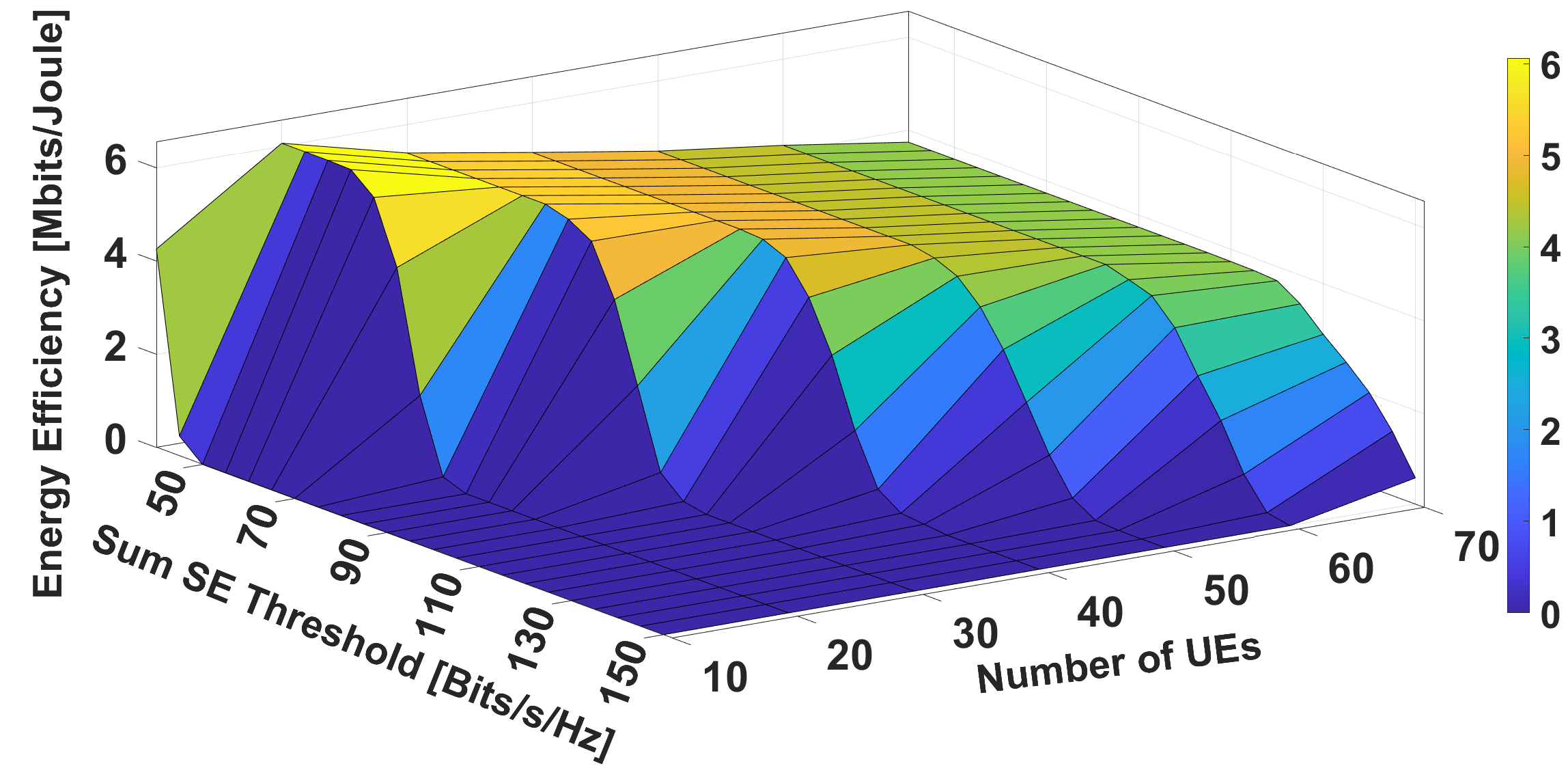}
    \caption{3D surface plot of EE as a function of the number of UEs and the sum SE threshold, when $M=100$.}
    \label{fig:EE_SE_UE}
\end{figure}
Fig.~\ref{fig:EE_SE_UE} illustrates the variation of EE with the number of UEs and the sum SE threshold. The results demonstrate a clear EE–SE trade-off similar to the plot in Fig. \ref{fig:EE_SE_AP}. For a fixed SE threshold, increasing the number of UEs initially enhances EE as more users contribute to system throughput, improving spatial diversity and AP utilization efficiency. However, beyond this point, EE begins to decline gradually as the number of UEs increases further. This drop occurs because supporting additional UEs requires higher transmit and processing power, leading to faster growth in total energy consumption compared to SE gain. For lower SE thresholds (up to 60~bits/s/Hz), a smaller number of UEs can meet the SE requirement efficiently, yielding a peak EE of around 5.6–6~Mbits/Joule at 20 UEs. As the SE threshold increases (upto 80~bits/s/Hz), the EE peak shifts to 30 UEs, with a slightly lower maximum EE of 5–5.3~Mbits/Joule. This shifting behavior continues for higher SE thresholds, indicating that larger UE density are needed to sustain higher throughput targets. 

Fig.~\ref{fig:EE_SE_AP} and Fig.~\ref{fig:EE_SE_UE} together highlight the importance of adaptive infrastructure design in distributed MIMO networks. They demonstrate how the interplay between transmit power, AP activation and AP–UE association governs the achievable balance among EE and SE. From a deployment perspective, these results emphasize that there exists an optimal network scale, in terms of both the number of APs and active UEs, that maximizes EE for a given SE requirement.

\begin{figure}[!h]
    \centering
    \includegraphics[width=0.60\textwidth]{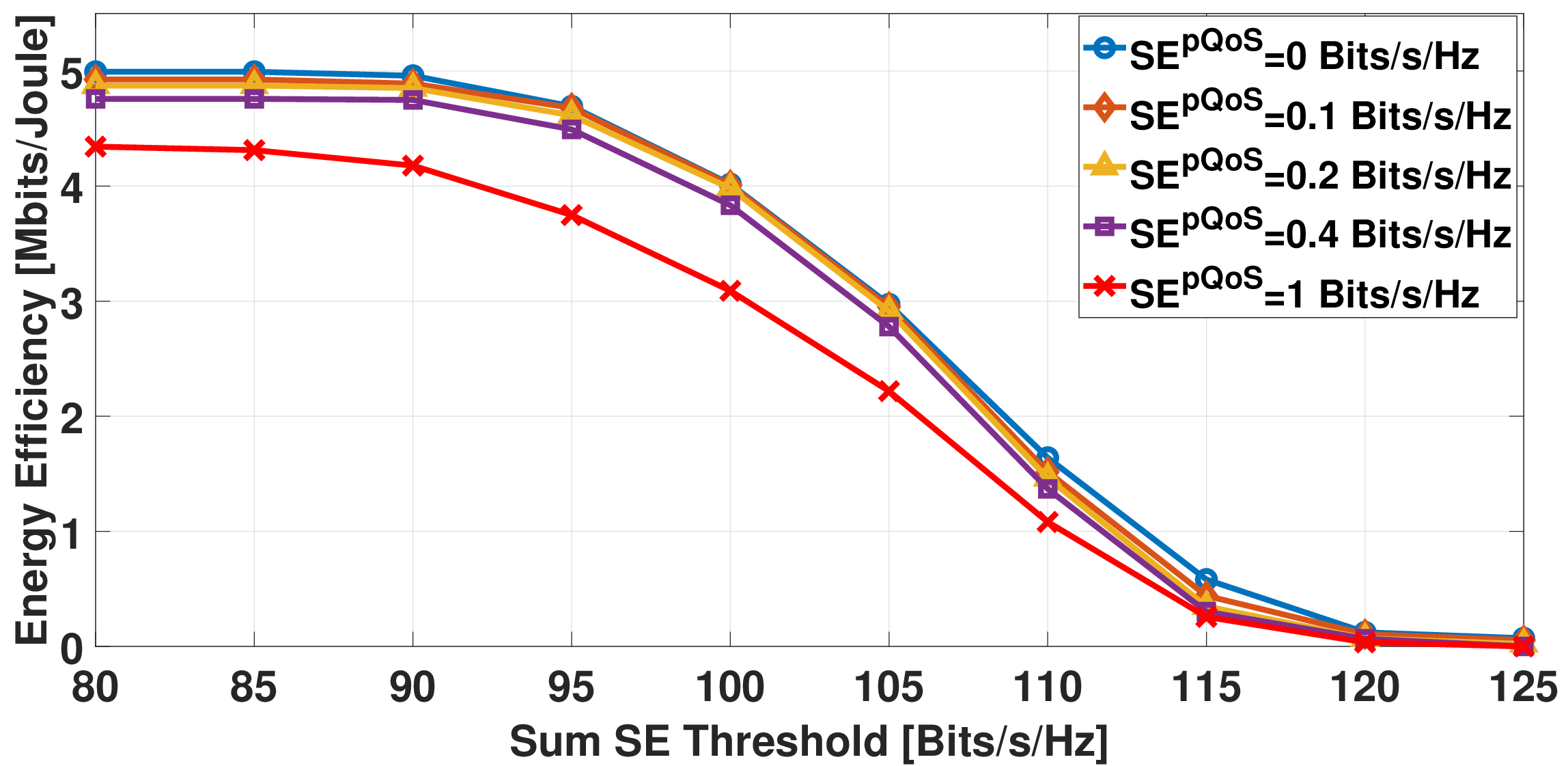}
    \caption{Energy efficiency versus sum SE threshold for different per-user SE thresholds ($\text{SE}^{\text{pQoS}}$), when $T=40$ and $M=100$}
    \label{fig:EE_SE_peruserSE}
\end{figure}
Fig.~\ref{fig:EE_SE_peruserSE} illustrates the impact of varying per-user SE threshold on the EE as the sum SE threshold increases. As per-user SE threshold increases, EE shows a gradual decrease across all sum SE levels, since enforcing higher individual user requirements limits overall spectral efficiency. However, this reduction remains minimal due to the efficient resource allocation achieved by the proposed joint optimization. When the per-user SE threshold increases from 0 to 0.4~bits/s/Hz, the EE drop is negligible, and even for the highest threshold, the maximum reduction is only about 0.6~Mbits/Joule. This demonstrates that the proposed method effectively balances user-level SE guarantees with energy-efficient system operation. The joint optimization framework maintains high EE while satisfying per-user QoS demands, highlighting its adaptability in achieving both performance and sustainability objectives in D-mMIMO networks.

Overall, the numerical results demonstrate that energy-efficient uplink operation in D-mMIMO systems critically depends on the joint optimization of AP activation, AP-UE association, and power control. The proposed framework is most beneficial under moderate traffic loads and intermediate EE
requirements, where infrastructure-related power consumption dominates and flexible adaptation yields substantial energy savings. In contrast, under highly stringent throughput constraints or extremely dense deployments, the achievable EE becomes increasingly limited by unavoidable circuit and processing power, causing performance differences among schemes to diminish. However the performance of the proposed joint framework schemes remain dominant. These observations highlight the importance of realistic power modeling and load-aware infrastructure adaptation when designing practical uplink D-mMIMO networks.

\section{Conclusion and Future Work}
\label{conclusion}
This paper investigated the fundamental EE–SE trade-off in D-mMIMO systems, emphasizing the joint optimization of AP activation, AP–UE association, and transmit power. The proposed framework provides a unified approach that captures both spectral performance and energy sustainability, addressing limitations of conventional schemes that optimize either in isolation. The performance analysis demonstrated that adaptive infrastructure control through selective AP activation and dynamic AP–UE association plays a decisive role in achieving energy-efficient operation while maintaining high spectral performance. The proposed joint optimization achieves a balanced trade-off across varying system loads, AP densities, and per-user SE constraints, effectively scaling with network size and user demand. From a deployment perspective, the results highlight that optimal network performance is achieved when system resources are adaptively coordinated rather than fully utilized, underscoring the importance of  context-aware optimization in next-generation wireless systems. Overall, the proposed framework establishes a strong foundation for energy-aware D-mMIMO networks, offering valuable insights for the sustainable evolution of 5G and beyond.

As future work, incorporating modulation and coding schemes into the optimization framework would allow a more refined characterization of the EE-SE trade-off. Moreover, extending the model to account for dynamic traffic patterns and time-varying user activity could further enhance the applicability of the proposed approach to real-time energy-efficient network control.

\appendix
\label{apx}

\subsection{Proof of Equivalence for Objective Function Maximization}
\label{apx_2}
Let the objective functions in \eqref{eq_4} and \eqref{eq_5} be defined as $H(\textbf{x}) = \frac{F(\textbf{x})}{G(\textbf{x}) + CF(\textbf{x})}$  and $H_1(\textbf{x}) = \frac{F(\textbf{x})}{G(\textbf{x})}$ respectively, with $F(\textbf{x}) > 0$, $G(\textbf{x}) > 0$, and constant $C > 0$. Expressing $H(\textbf{x})$ in terms of $H_1(\textbf{x})$ yields:
\begin{align*}
H(\textbf{x}) = \frac{H_1(\textbf{x})}{1 + CH_1(\textbf{x})}.
\end{align*}
The derivative of $H(\textbf{x})$ with respect to $H_1(\textbf{x})$ is:
\begin{align*}
\frac{dH}{dH_1} = \frac{1}{(1 + CH_1)^2} > 0,
\end{align*}
which is strictly positive for all $H_1(\textbf{x}) > 0$. Therefore, $H(\textbf{x})$ is a strictly increasing function of $H_1(\textbf{x})$, confirming that maximizing $H_1(\textbf{x})$ is equivalent to maximizing $H(\textbf{x})$. This proof establishes the equivalence between optimization problems \eqref{eq_4_m} and \eqref{eq_5_m}.

\subsection{Proof of Lemma \ref{lem_1}}
\label{apx_lemma1}
The uplink SINR of user $t$ is given by $\Gamma_t = \mathrm{DS}_t / \mathrm{I}_t$, where $\mathrm{DS}_t \ge 0$ and $\mathrm{I}_t > 0$. Consider the quadratic function $\phi(z_t) = 2 z_t \sqrt{\mathrm{DS}_t} - z_t^2 \mathrm{I}_t$, which is concave in $z_t$. Maximizing $\phi(z_t)$ with respect to $z_t$ yields the optimal auxiliary variable $z_t^{\star} = \sqrt{\mathrm{DS}_t / \mathrm{I}_t}$, at which
$\phi(z_t^{\star}) = \mathrm{DS}_t / \mathrm{I}_t = \Gamma_t$. Hence, $\phi(z_t) \le \Gamma_t$ for all $z_t$, with equality at $z_t = z_t^{\star}$, establishing that the quadratic expression in~\eqref{eq_11} provides a concave lower bound on $\Gamma_t$ that is tight at the optimum.

\subsection{Proof of Lemma \ref{lem_2}}
\label{apx_lemma2}
For $u \ge 0$ and $v>0$, define $\phi(b)=2b\sqrt{u}-b^{2}v$, which is a concave quadratic function in $b$. Its maximum is attained at $b^{*}=\sqrt{u/v}$, yielding $\phi(b^{\star})=u/v$. Hence, $\phi(b)\le u/v$ for all $b$, with equality at $b=b^{*}$. This shows that the quadratic expression in~(25) is a concave lower bound on $u/v$ and is tight at the optimum, completing the proof.

\subsection{Proof of Theorem \ref{t_1}}
\label{apx_3}
\begin{itemize}
 \item \textbf{Feasibility Preservation:} By Lemma~\ref{lem_2}, the quadratic transformation $2b^*\sqrt{u} - (b^*)^2v$ yields a tight concave lower bound for the fractional objective $\frac{u}{v}$, which becomes exact at \( b = b^* \). This ensures equivalence between the transformed problem \eqref{eq_14_m} and problem \eqref{eq_8_m}. Moreover, the quadratic transformation of the SINR term $\Gamma_t$ (as established in Lemma~\ref{lem_1}) ensures that the surrogate constraint \eqref{eq_11} lower bounds $\Gamma_t$, thereby satisfying the original SINR constraint \eqref{eq_4_3}. Thus, under the setting \( z_t = z_t^* \), and following Remark 1, solving the approximated problem \eqref{eq_8_m} guarantees equivalence with the original problem \eqref{eq_4_m}.
 \item \textbf{Monotonic Improvement:} To demonstrate the non-decreasing nature of the function  $f(\bm{\eta^u},\textbf{D},\textbf{s})= 2b\sqrt{u} - b^{2}v$, assume that ${\bm{\eta^u}}^*$ represents the optimal value of $f$, when $\textbf{D}$ and $\textbf{s}$ are fixed. Given this, the inequality $f({\bm{\eta^u}}^*,{\textbf{D}}^{(i)},{\textbf{s}}^{(i)}) \geq f({\bm{\eta^u}}^{(i)},{\textbf{D}}^{(i)},{\textbf{s}}^{(i)})$ always holds due to the  concavity of the function $f$ with respect to $\bm{\eta^u}$. When optimizing $\textbf{D}$ to $\textbf{D}^*$ and $\textbf{s}$ to $\textbf{s}^*$, with $\bm{\eta^u}$  fixed at ${\bm{\eta^u}}^{*}$,  the inequality $f({\bm{\eta^u}}^{*},{\textbf{D}}^{*},{\textbf{s}}^{*}) \geq f({\bm{\eta^u}}^{*},{\textbf{D}}^{(i)},{\textbf{s}}^{i})$ always holds as $f$ is concave with respect to $\textbf{D}$ and $\textbf{s}$. Therefore, combining these observations, we see that  $f({\bm{\eta^u}}^{(i+1)},{\textbf{D}}^{(i+1)},{\textbf{s}}^{(i+1)}) \geq f({\bm{\eta^u}}^{(i)},{\textbf{D}}^{(i)},{\textbf{s}}^{(i)})$, indicating $f$ is  non-decreasing at each iteration. This non-decreasing trend makes the optimization function monotonically increasing in each iteration and also the optimization function is bounded from above, ensuring the convergence of the optimization algorithm (Algorithm~\ref{Algorithm 1}), as the function does not increase indefinitely but plateaus at the maximum value.
 \item \textbf{Stationary Point:} From the monotonic improvement and feasibility preservation, solving the transformed problem \eqref{eq_14_m} via alternating optimization of subproblems \eqref{eq_15_m} and \eqref{eq_16_m} leads to a stationary point $\{{\bm{\eta^u}}^*, \textbf{D}^*, \textbf{s}^* \}$. Given the equivalence of the transformed problem to the original problem, this stationary point also satisfies the original problem \eqref{eq_4_m}.
\end{itemize}


\end{document}